\documentclass[twocolumn]{aastex631}
\usepackage{{booktabs}}
\usepackage{longtable}
\usepackage{rotating, graphicx}
\usepackage{hyperref}

\newcommand\noclusters{35}
\newcommand\fb{$f_{\rm b,q}$}

\newcommand\MLMSfb{$f_{\rm b,q,M}$}
\newcommand\MLMSfbrc{$f_{\rm b,q, M,rc}$}

\begin{document}

\title{Stellar Dynamics in Open Clusters Increases the Binary Fraction and Mass Ratios: Evidence from Photometric Binaries in 35 Open Clusters}

\author[0000-0002-9343-8612]{Anna C. Childs}
\affiliation{Center for Interdisciplinary Exploration and Research in Astrophysics (CIERA) and Department of Physics and Astronomy Northwestern University,
1800 Sherman Ave, Evanston, IL 60201 USA}

\author[0000-0002-3881-9332]{Aaron M. Geller}
\affiliation{Center for Interdisciplinary Exploration and Research in Astrophysics (CIERA) and Department of Physics and Astronomy Northwestern University,
1800 Sherman Ave, Evanston, IL 60201 USA}



\begin{abstract}
Using the Bayesian Analysis of Stellar Evolution-9 (BASE-9) code and Gaia DR3, Pan-STARRS, and 2MASS data, we identify  photometric binaries  in \noclusters \, open clusters (OCs) and constrain their masses.  We find a strong correlation between the binary fraction and cluster dynamical age and an even stronger correlation between core binary fraction and cluster dynamical age.  We find the binary mass-ratio ($q$) distribution of dynamically young OCs is statistically distinct from that of the old OCs.  On average, dynamically young OCs display multi-modal $q$ distributions rising toward unity and toward our detection limit of $q=0.5$ while more dynamically evolved clusters display more uniform $q$ distributions often with a peak near $q=1$.   Interestingly, the uniform $q$ distribution with a peak near $q=1$ is consistent with binaries in the field.  We also observe a similar transition from multi-modal to unimodal $q$ distributions when comparing low mass to high mass OCs in our sample.  Lastly, we find a correlation between the median $q$ of the binary population in a cluster and the cluster dynamical age.  We interpret these results as an indication that dynamical encounters tend to increase the fraction of high-mass-ratio binaries within a given cluster -- particularly within the cluster's core where stellar dynamics are likely more important.  This may be the result of stellar exchanges that tend to produce binaries with larger $q$ and/or the preferential disruption or evaporation of lower $q$ binaries. 

\end{abstract}



\section{Introduction} \label{sec:intro}
Most Sun-like stars in the galaxy are in binary or higher-order systems, and the binary occurrence rate increases with primary stellar mass \citep{Raghavan2010,san12,Duchene2013,cab14}.  Approximately 30\% of M dwarfs in the solar neighborhood have a companion \citep{Winters2019} and $\sim10–20\%$ of Sun-like stars are hierarchical triples or higher-order multiples \citep{Tokovinin2014, Raghavan2010}. While binary and higher ordered systems are ubiquitous, their formation channels and evolution are still not fully understood (for a comprehensive overview of multiple star formation mechanisms, dynamical evolution, and the role of environmental conditions, see \citealt{Offner2023}). 

Open clusters (OCs) are thought to be the birth places of most stars with masses $M \geq 0.5 \, M_{\odot}$ which eventually disperse to populate the galactic field \citep{lad03, eva09, bre10}.   Because the stars in a given OC have the same age and metallicity, and OCs in our galaxy sample a wide range in age, metallicity and density, OCs are ideal environments to learn about the birth characteristics and evolutionary histories of binary stars.  

Within star clusters, various processes of stellar dynamics and relaxation work together to shape the stellar and binary populations.  One primary driver is mass segregation, the theoretical prediction that more massive cluster members will sink towards the center of the cluster more rapidly.  This effect is most pronounced for more massive single stars and binary stars, while less massive single stars remain further from the core and are preferentially lost from the OC \citep{Cote1991, Layden_1999, Fregeau_2002, Childs2024, Motherway2024, Zwicker2024}. Theoretical expectations predict that the effects of mass segregation and cluster evaporation lead to the global binary fraction of an OC increasing with time, as lower-mass single stars are preferentially lost from the cluster \citep{delaFuente1998, Sollima2007, Fregeau2009}.

Close stellar dynamical encounters can also play a key role, when stellar densities are high enough. 
In addition to altering the overall binary fraction of the cluster, dynamical interactions can also modify the masses and orbital properties of the binaries.  For instance, if the energy of an incoming star is larger than the binding energy of the binary in the encounter, the binary is expected to be disrupted.  Thus, one often conceptualizes a limit on the binary separation, where widely separated binaries (soft binaries) will not survive a typical stellar encounter, and the remaining binary population will thus comprise more closely separated binaries (hard binaries) \citep{heggie75, Geller_2013}.  A similar argument can be made for binary secondary mass (or mass ratio) with all else held constant, where low-mass-ratio binaries are more easy to disrupt.  Furthermore, previous models and simulations of binary formation suggest that wider binaries tend to form with lower mass ratios \citep{Bate2000, Bate2009}.  If the encounter rate is high enough, these destructive encounters may be expected to reduce the binary fraction in a cluster, and preferentially remove low-mass-ratio binaries. 

If a binary survives a strong encounter, conventional wisdom is that the binary is more likely to contain the most massive stars in the encounter.  For instance, if the binary begins with a low-mass secondary star, and an incoming star has a larger mass than this secondary, this incoming star is likely to be exchanged into the binary.  These exchange encounters are thus expected to increase the binary mass ratios over time \citep{marks2011, Geller_2013}.  

In highly dynamically active environments, when at least three single stars are found close together, binary creation can also occur \citep{Mansbach1970, Ginat2024}.  This however, is highly unlikely in open clusters because the densities are too low.

Because of these processes, the binary population in an OC will change over time, both globally and as a function of radius from the cluster center.  Thus, it is not clear how much of current observations of both field and cluster binaries are representative of primordial binary populations or the dynamical history of such. More detailed studies of binary populations in OCs are needed to help disentangle the primordial and dynamically evolved populations in the observations to better understand stellar formation and evolution. 

There have been many observational campaigns aimed at testing the aforementioned theoretical predictions, each with their own limitations.  One of the most promising methods utilizes new high-precision, wide-field photometric surveys and offers a path toward a more complete identification of binaries in OCs, than is possible by e.g., time-series radial-velocity or photometric surveys.  As compared to a single star of a given mass, the photometry of a binary with the same primary mass will be brighter (a result of the combined magnitude of two stars) and redder (a result of the combined color from two different massed stars) \citep[e.g.][]{Hurley1998}.  This difference in brightness and color allows binaries to be identified on a color magnitude diagram (CMD) from a single-epoch of photometry (in at least two different filters).  Other notable methods for identifying unresolved binaries in OCs include using a mixture model of single and binary stars to constrain the binary properties of a given OC \citep{Li2020}, ``ridgeline" isochrone fitting \citep{Jiang2024}, and photometric multi-band fitting techniques \citep{Liu2025, Liu2025b}.

The photometric binary fraction in a limited sample of OCs \citep{Frinchaboy2015, Cohen2020, Cordoni2023, Jadhav2021, Donada2023, Pang2023, Childs2024, Motherway2024} and  globular clusters (GCs) \citep{Sollima2007,Milone2012, Ji2015} have been investigated in previous studies, although findings and conclusions differ from study to study.  \citet{Donada2023}, \citet{Cordoni2023} and \citet{Childs2024} find a correlation between binary fraction and cluster age in their samples of OCs, which might indicate that the processes of mass segregation and cluster evaporation dominate over binary disruption.  However \citet{Pang2023} do not observe this trend in their OC data.  \cite{Cordoni2023} also found a positive correlation between binary fraction with both central density and total mass.   \cite{Childs2024} did not observe this trend, though they had a much more limited number of OCs in their sample.

Previous studies showed that the binary fraction in the cores of GCs decreases with cluster age \citep{Sollima2007} and cluster dynamical age \citep{Ji2015}, the number of half-mass relaxation timescales the cluster has lived through, which might be expected if binary destruction is dominant.  More recently, \cite{Mohandasan2024} constrained the photometric binary fraction in 14 Magellanic Cloud stellar clusters, 67 Galactic GCs, and 78 OCs.  They found no evidence of a correlation between binary fraction and cluster age or dynamical age.  However, similar to \cite{Ji2015} and \cite{Milone2012}, they did find a strong anti-correlation between binary fraction in the core and total cluster mass, although binary fractions can vary widely for clusters of similar total mass.  This might indicate that both dynamics and variation in primordial binary fraction play a role in observed cluster binary fractions.

Photometric studies of OCs have also investigated the cluster mass-ratio ($q$) distributions.  When considering the OCs Alpha Persei, Praesepe, and NGC 1039, \citet{Malofeeva2023} found that the $q$ distribution for the unresolved binaries could be approximated by a Gaussian with a wide range of maximum between 0.22 and 0.52. \cite{Albrow2024} considered the $q$ distributions in the Hyades and Praesepe clusters finding that higher $q$ binaries are more common than lower $q$ binaries.  However, \cite{Alexander2025} found the $q$ distribution is fairly flat for the six young OCs they considered.  \cite{Childs2024} also found a flat $q$ distribution for six OCs that covered a wide range of ages.

In summary, previous studies of photometric binary fractions are sometimes inconclusive or at odds with existing literature.  This is likely because determining the photometric binary fraction is very dependent on the quality of the data, the accuracy of the isochrone fit to the data, the number and wavelength coverage of photometric filters included, and the analysis method used.  Most, and sometimes all, of these factors vary between different studies in the literature.  Furthermore, previous studies often employ a ``chi-by-eye" fit, where the isochrone that visually fits the data best is used.  This is far from ideal, can easily suffer from human bias, and is typically only feasible using one CMD (e.g., two or possibly three filters). 

To help address these difficulties and inconsistencies, we employ a Bayesian technique to identify a posterior distribution for the cluster parameters that defines a family of isochrones that describe the observational data \citep{Cohen2020, Childs2024, Motherway2024, Zwicker2024}. More specifically, we use the Bayesian Analysis of Stellar Evolution with Nine Parameters (BASE-9) code, which yields precise likelihoods on the cluster parameters, binarity and mass of any given star in the cluster \citep{vonHippel2006, vanDyke2009, Robinson2016}, to analyze Gaia DR3, Pan-STARRS and 2MASS photometry for a sample of \noclusters\, nearby low-redenning OCs. 
BASE-9 considers all filters from all three surveys (a total of 11 filters) simultaneously when identifying the photometric binaries.   This combination of state-of-the-art software and more recent and precise data, allows us to determine more accurate and self-consistent cluster parameters and binary fractions than was previously possible.

In Section \ref{sec:methods} we discuss our observational data and methods for determining our final sample of photometric binaries.  We then analyze the BASE-9 results for each of the \noclusters\ OCs and look for trends with cluster parameters that yield insight into binary formation and evolution.  In Section \ref{sec:Results} we present these results, which we discuss further in Section \ref{sec:Discussion}.  Finally, we conclude with a summary of our study in Section \ref{sec:Conclusions}.

\section{Observational Data} \label{sec:methods}
We follow the methods of \cite{Childs2024}-henceforth referred to as Paper I, to determine cluster members, cluster parameters, and stellar mass and binarity on a star-by-star basis for the cluster members.  To do this, we use Gaia DR3, 2MASS, and Pan-STARRS data in tandem with BASE-9.  Although our overall data analysis procedure can be found in Paper I and references there in,  here we make some notable improvements to our previous methods which we discuss below.



\subsection{Cluster Sample}\label{subsection:cluster_sample}

Our sample includes all OCs from \cite{Hunt2023} that have a declination $\delta > 30^{\circ}$, a distance between 500 and 4000 pc, reddening $A_{\rm v}<1$, an age greater than $10 \, \rm Myr$, and at least 500 members. This includes all OCs in Paper I with the exception of NGC 6791 which is beyond 4000 pc.  To constrain the cluster core and tidal radii, we apply King fits to our final cluster members (more details on this in Section \ref{sec:cluster_params}), and we ensure we have data out to at least three core radii for each OC.  We only consider OCs with $\delta > 30^{\circ}$ as this is the region for which Pan-STARRS1 photometry is available.  We choose the distance cuts because ($i$) OCs closer than $\sim500 \, \rm pc$ have large angular sizes, often resulting in lots of field star contamination and ($ii$)  OCs beyond $\sim4000 \, \rm pc$ become too faint.  Our age cut is chosen to exclude the youngest clusters where enough gas may still be present to create significant differential reddening in the cluster, and also because the models become less reliable for such young clusters.  While there exist many small OCs within these distance and age cuts, we only look at the OCs with at least 500 members in the \citet{Hunt2023} catalog.  These criteria yield a sample of 37 OCs.  Of these 37 OCs, we remove two that we find to be too overwhelmed by the field-star population to yield reliable results from our analysis method (NGC 2324 and NGC 6716), thus bringing our final sample of OCs to \noclusters.  We list these clusters in Table \ref{tab:cluster_params}.

 \subsection{Determining Cluster Membership}
To first separate the likely cluster members from the field stars, we query all the Gaia DR3 data within a large radius around the cluster center (center location is taken from previous literature).  We then look at the distribution of Gaia DR3 radial-velocity, parallax, and proper-motion measurements for peaks that would correlate with the cluster members.  While this method worked well for the six clusters considered in Paper I, it failed for many of the clusters in our extended sample, specifically for clusters that lie close to the Galactic plane and/or have substantial field star contamination.  To circumvent this issue we employ HDBSCAN, a hierarchical density based clustering algorithm that finds clusters of data in $n$-dimensional space \citep{McInnes2017}.  Using HDBSCAN in conjunction with Gaia data to search for OCs has been done previously by multiple groups \citep{Hunt2021, Tarricq2022, Hunt2023, Qin2023, Alfonso2024} and has proven to be a very powerful tool for these types of studies. 

We apply HDBSCAN to data within $0.5-1.5^{\circ}$ from the cluster center (depending on the cluster's angular size) to look for groups of data points with similar right ascension, declination, parallax, and proper motion (in both dimensions).  HDBSCAN then returns multiple groups of data and the probability that each data point belongs to their respective group.  To find the HDBSCAN group that corresponds with the OC, we plot the CMD of the HDBSCAN group and compare the median RA and DEC of the group to previous literature values of the cluster's expected center.  Through this method we are able to locate OC members even if the peaks in their distributions of Gaia kinematic and parallax measurements are obscured by field-star contamination.  Once we identify the HDBSCAN group that corresponds to the OC, we fit Gaussian functions to the distributions of the Gaia radial-velocity, proper-motion and parallax measurements for the HDBSCAN group (similarly to our method in Paper 1).  Next, we apply these fits to the data set from a larger radial region on the sky (at least 3 core radii, see below) and calculate the $p$-values for each star (our estimate of a cluster membership probability), as defined in Paper 1.

\subsection{Quantifying Field Star Contamination}
\label{ssec:contamination}
Because the degree of field star contamination changes for each cluster, we cannot use the same cutoff in $p$-value in every OC to reliably determine members.  Instead, we quantify the amount of field star contamination within each dataset queried for a given cluster and find what minimum $p$-value will result in $\sim 1 \%$ field contamination -- the lowest percentage of field star contamination we find possible for all OCs in our sample.

To do this, we first only consider stars that have a $p$-value$>0.002$, which is approximately the three-sigma bound for a normal distribution.  Next, we count how many stars are within a small radial range from the cluster center (within 0.5 degrees or less, depending on the angular size of the OC), $N(r_{\rm in})$.  We expect minimal field star contamination in this sample, as this covers the core of the cluster (and only contains stars that pass our membership threshold).  Then we count how many stars are in the outermost annulus of the data queried for a given OC, with a width of 0.1 degree, $N(r_{\rm out})$.  We have queried Gaia DR3 data out to a large enough radius for all clusters that this outermost annulus should contain most, if not all, field stars.  If we then assume that this field star count is isotropic across the radial range of the OC, we can quantify what fraction of our identified cluster members are fields stars with \begin{equation}
    f_{\rm c}=1- \frac{N(r_{\rm in})}{ CN(r_{\rm out})+N(r_{\rm in})},
\end{equation}
where $C$ is a normalization constant equal to the area of the inner circle, where $N(r_{\rm in})$ is taken, over the area of the outer ring, where $N(r_{\rm out})$ is taken.

If the field star contamination is calculated to be greater than $1\%$ for a given OC using a $p$-value=0.002, we increase the $p$-value minimum and repeat this process until we find a $p$-value that results in $\sim 1 \%$ field star contamination.  The minimum $p$-value used for each cluster is listed in Table \ref{tab:cluster_params}.

\section{Analysis \& Results} \label{sec:Results}
In this section, we present our results from BASE-9 and describe our methods of analysis.  We discuss the cluster parameters, the cluster binary fractions and the binary mass-ratio distributions.  We also compare our results to previous literature where available.

\subsection{Cluster parameters}\label{sec:cluster_params}
Using the photometry from the likely members we identify using the method described above, we employ BASE-9, which uses Markov Chain-Monte Carlo sampling, to find a set of PARSEC isochrones \citep{Bressan2012} that describe the photometry in all 11 photometric filters simultaneously. We take the resulting median values of the BASE-9 posterior distributions for reddening ($A_{\mathrm V}$), distance, age, and metallicity as summary values for a given OC. Table \ref{tab:cluster_params} lists the BASE-9 derived cluster parameters (median values and $1\sigma$-equivalent uncertainties) for the \noclusters\,clusters we consider here.  We use the median posterior values to determine the ``median isochrone"  for visualization purposes.  

In Figure \ref{fig:CMDs} in the appendix, we show the CMDs of all the final cluster members as well as the median isochrone in red and color the stars according to our classification of the BASE-9 results: black if they are single stars, and the binary stars are colored by their mass ratio.  The cluster members found by \cite{Hunt2023} that lie within the radial extent we consider are shown with gray `X's.  BASE-9 will reject exotic members (blue stragglers, yellow stragglers, sub-subgiants, etc.) and in our implementation, will also reject white dwarfs and binaries containing white dwarfs that lie far from the main-sequence\footnote{Note that BASE-9 can be configured to also study white dwarfs, and to include them in the cluster member sample.  For our purposes, we do not utilize this feature.} as photometric non-members, and this is where we find the biggest disagreement between ours and the \cite{Hunt2023} membership selections.

Note that BASE-9 uses the full posterior distribution when determining star-by-star masses and memberships.  We only consider iterations that have a membership value of ``True" in the posterior distribution of a given star.  Similar to \cite{Cohen2020}, we consider a star a binary if the median value of the posterior distribution in secondary mass is $\geq 3\sigma$ from zero. The mass ratios and masses we report for a given star are the median values of the posterior distributions for each parameter.

As in Paper I, we find that the median isochrone fits the blue edge of the main-sequence track nicely in most clusters, but BASE-9 often has difficulty fitting the turn off, the giant branch, and constraining the binary mass ratios near the `blue hook' morphology in clusters where this is present.  It is not clear if this is a limitation of the models, or BASE-9's method, or the precision of the photometry, but we do note that modeling the evolutionary tracks of more evolved stars is difficult to do \citep{Martins2013}.  Therefore we limit our analysis of the binary stars to the main-sequence track in this paper.

\begin{figure}[hbt!] 
\includegraphics[width=1\columnwidth]{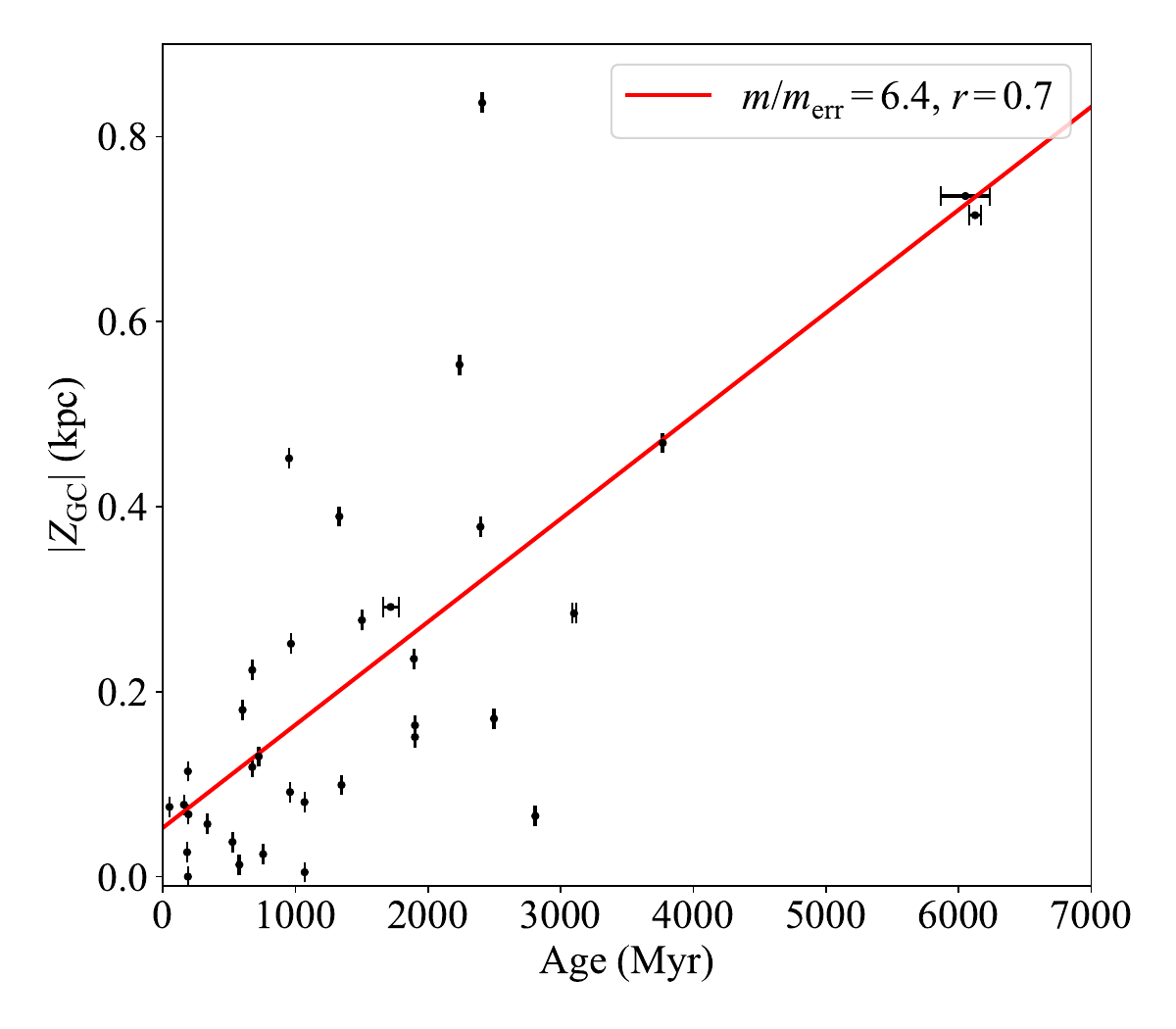}
    \caption{OC age versus OC distance above (or below) the galactic plane, $|Z_{\rm GC}|$.} 
    \label{fig:gal_z}
\end{figure}

With our BASE-9 derived ages we recover the known correlation between cluster age and distance above (or below) the galactic plane, $|Z_{\rm GC}|$ \citep[e.g][]{Janes1994, Friel1995}.  Figure \ref{fig:gal_z} shows OC age versus $|Z_{\rm GC}|$ with the line of best fit in red.  The linear fit has a Pearson correlation coefficient of $r=0.7$ and a $t-\rm statistic$, the slope divided by the standard error of the slop, of $m/m_{\rm err}=6.4$.

\begin{figure}[hbt!] 
\includegraphics[width=1\columnwidth]{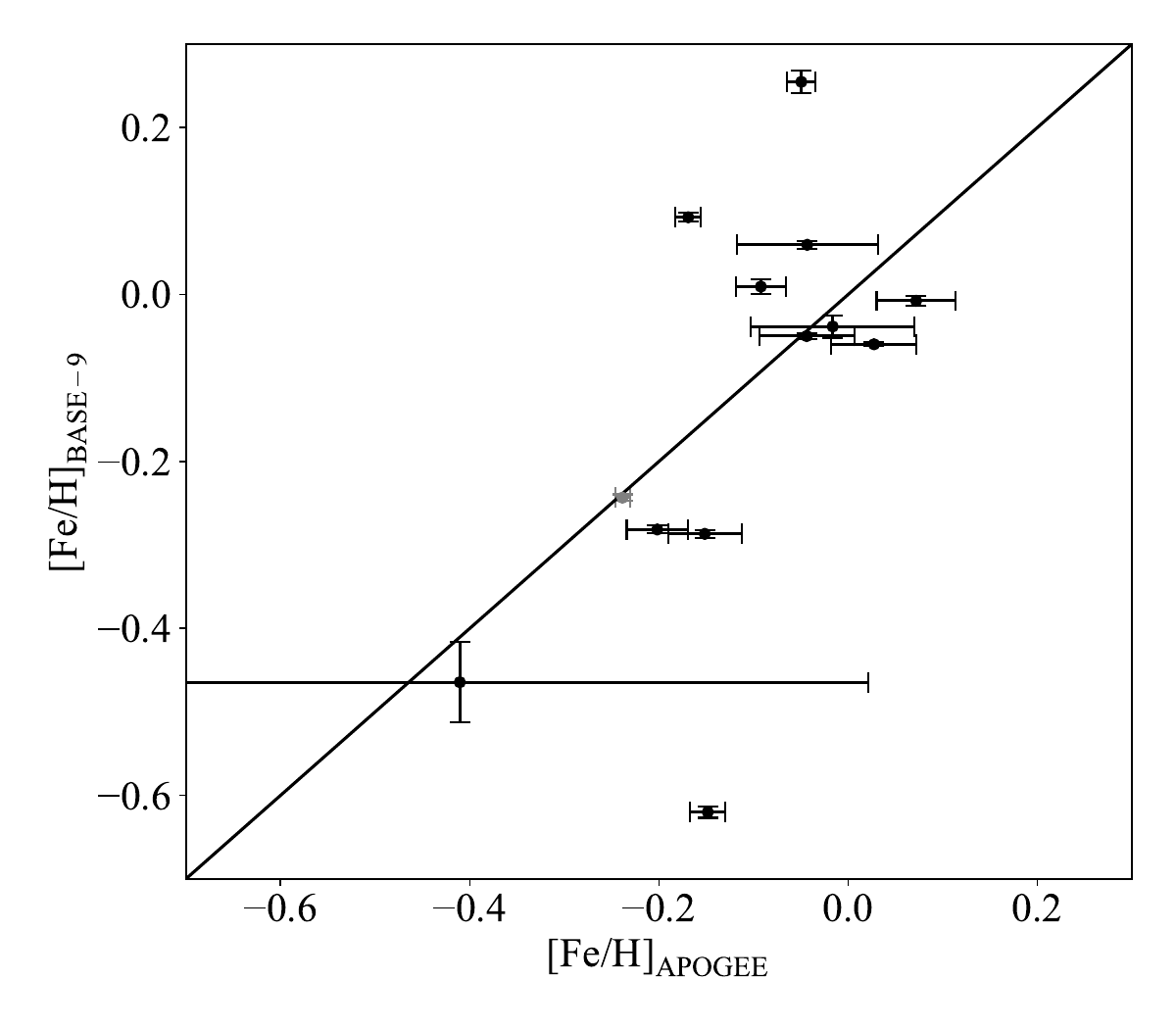}
    \caption{Average values of the cluster member metallicity from APOGEE, where available, versus the BASE-9 derived cluster metallicity. For the black points, the APOGEE error bars show the standard deviation of the multiple measurements taken in the given cluster, and the gray point identifies the single OC where only one measurement was available and so instead, we use the APOGEE uncertainty on the measurement for the error bar.  We include the 1:1 line in black.} 
    \label{fig:APOGEE}
\end{figure}

To check how well BASE-9 is able to recover the cluster metallicity, we compare the BASE-9 median cluster values to the metallicity of the cluster members from the spectroscopic APOGEE survey \citep{Majewski_2017}. After matching the Gaia source ids of our identified cluster members to the APOGEE catalog we find APOGEE spectroscopic data for at least one cluster member in 13 of the clusters in our sample.  In Figure \ref{fig:APOGEE} we show the average values of the cluster member metallicity from APOGEE, where available, versus the BASE-9 derived cluster metallicity. 
We find overall good agreement and a symmetric scatter of the BASE-9 derived [Fe/H] and the APOGEE [Fe/H], about the 1:1 line and suggest that some of the discrepancy can be attributed to two main features.  First, BASE-9 samples over all cluster members whereas the APOGEE metallicities are from only a few cluster members.  Second, photometry is not as sensitive to metallicity as spectra, and so our BASE-9 constraints from photometry are likely not as precise as APOGEE spectra measurements.  We note that the internal metallicity spread for a given OC is on the order of  $~0.2 \, \rm dex$ \citep{Heiter2014} which is consistent with the discrepancy between most of the measurements shown here.  Thus we conclude that BASE-9 is recovering reasonable metallicities.

\begin{figure*}
\includegraphics[width=0.5\textwidth]{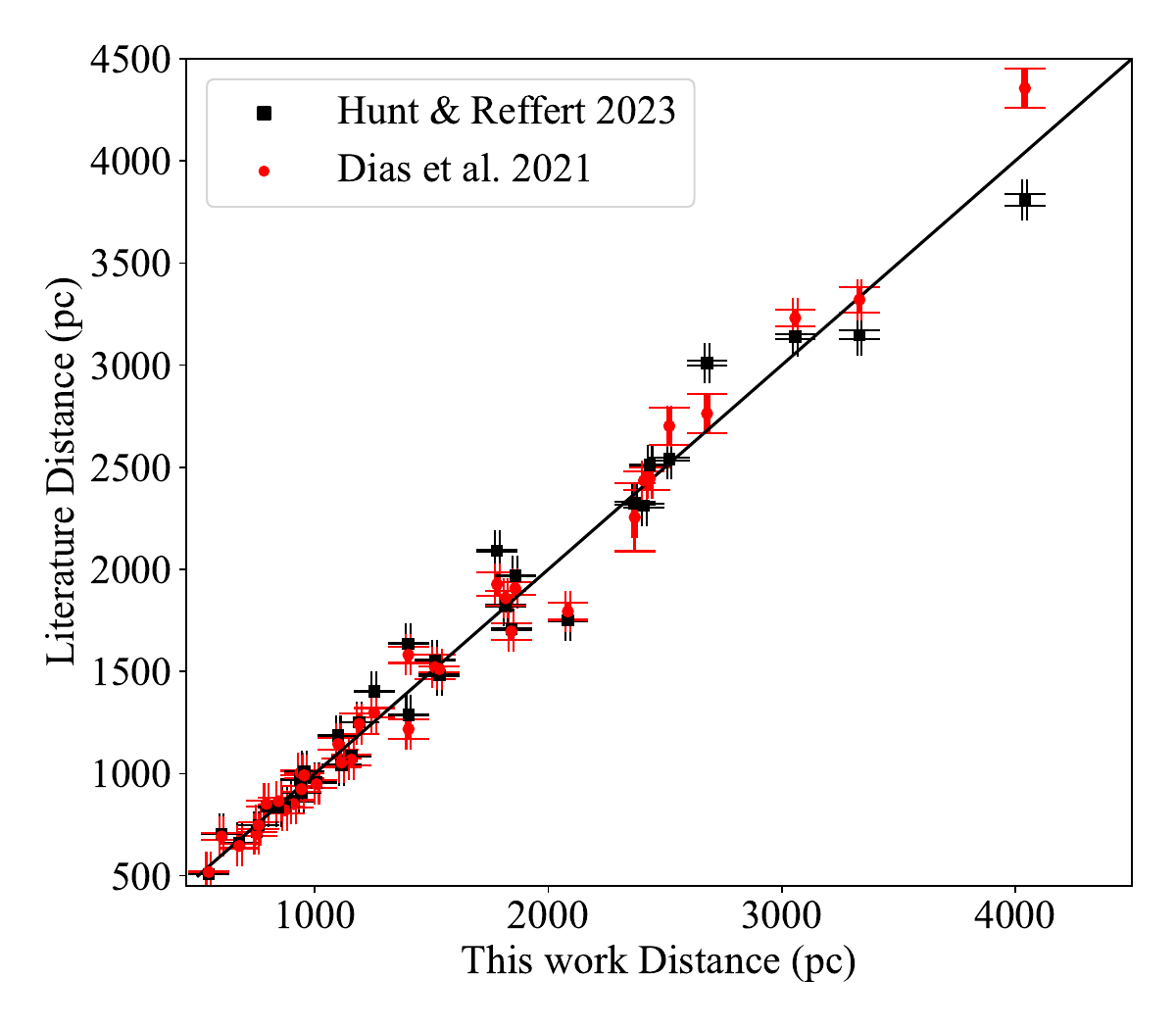}
\includegraphics[width=0.5\textwidth]{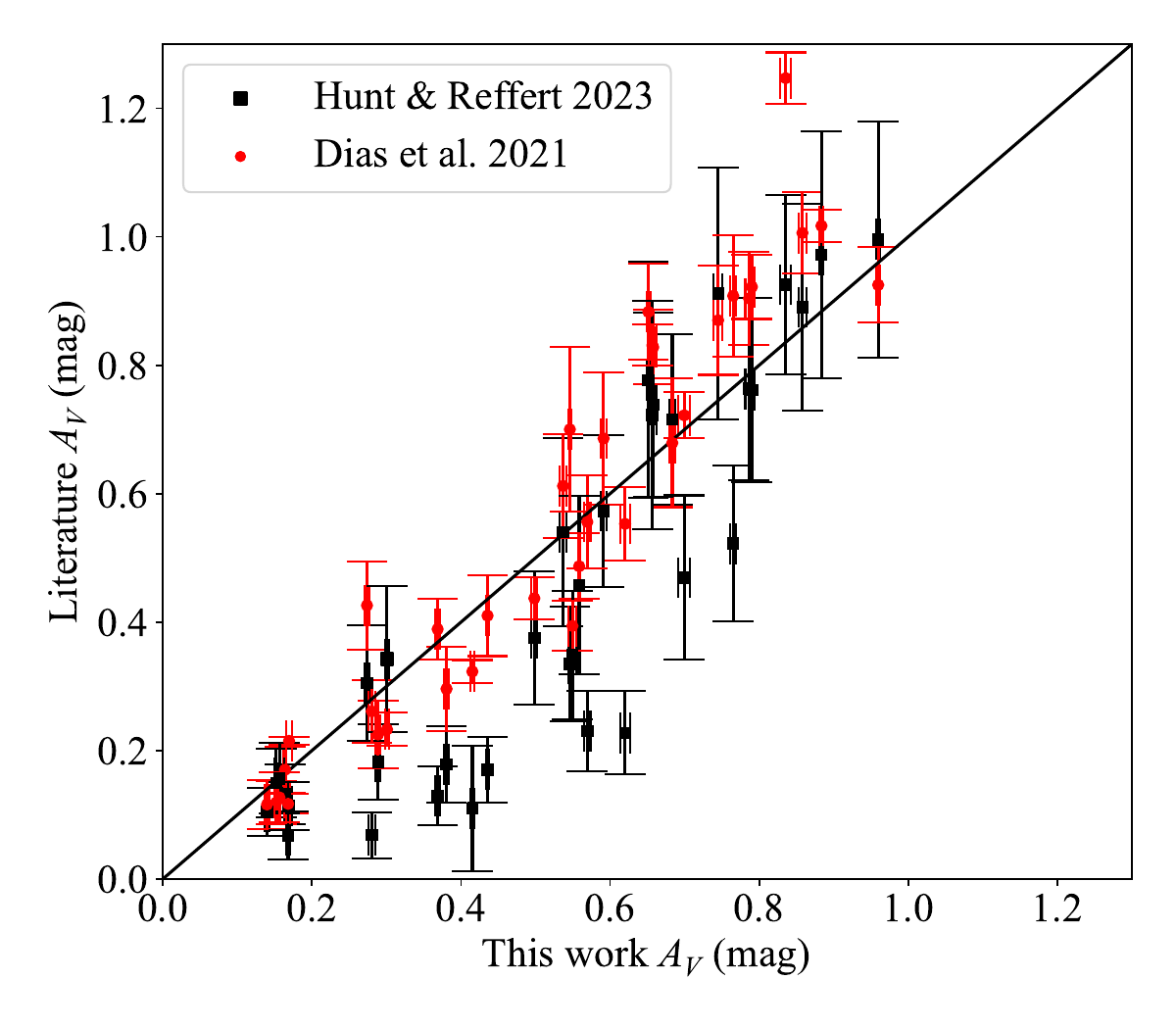}
\includegraphics[width=0.5\textwidth]{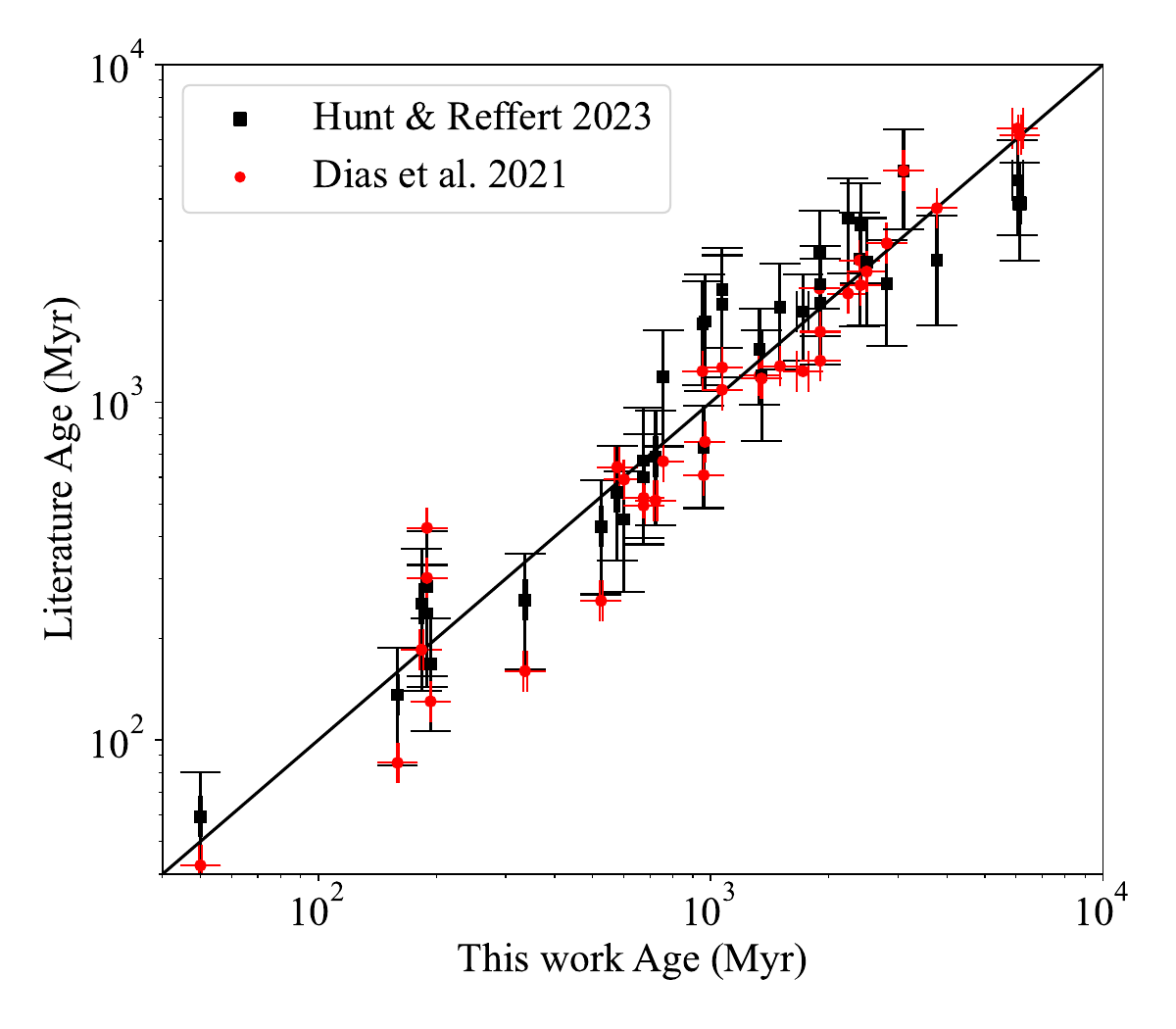}
\includegraphics[width=0.5\textwidth]{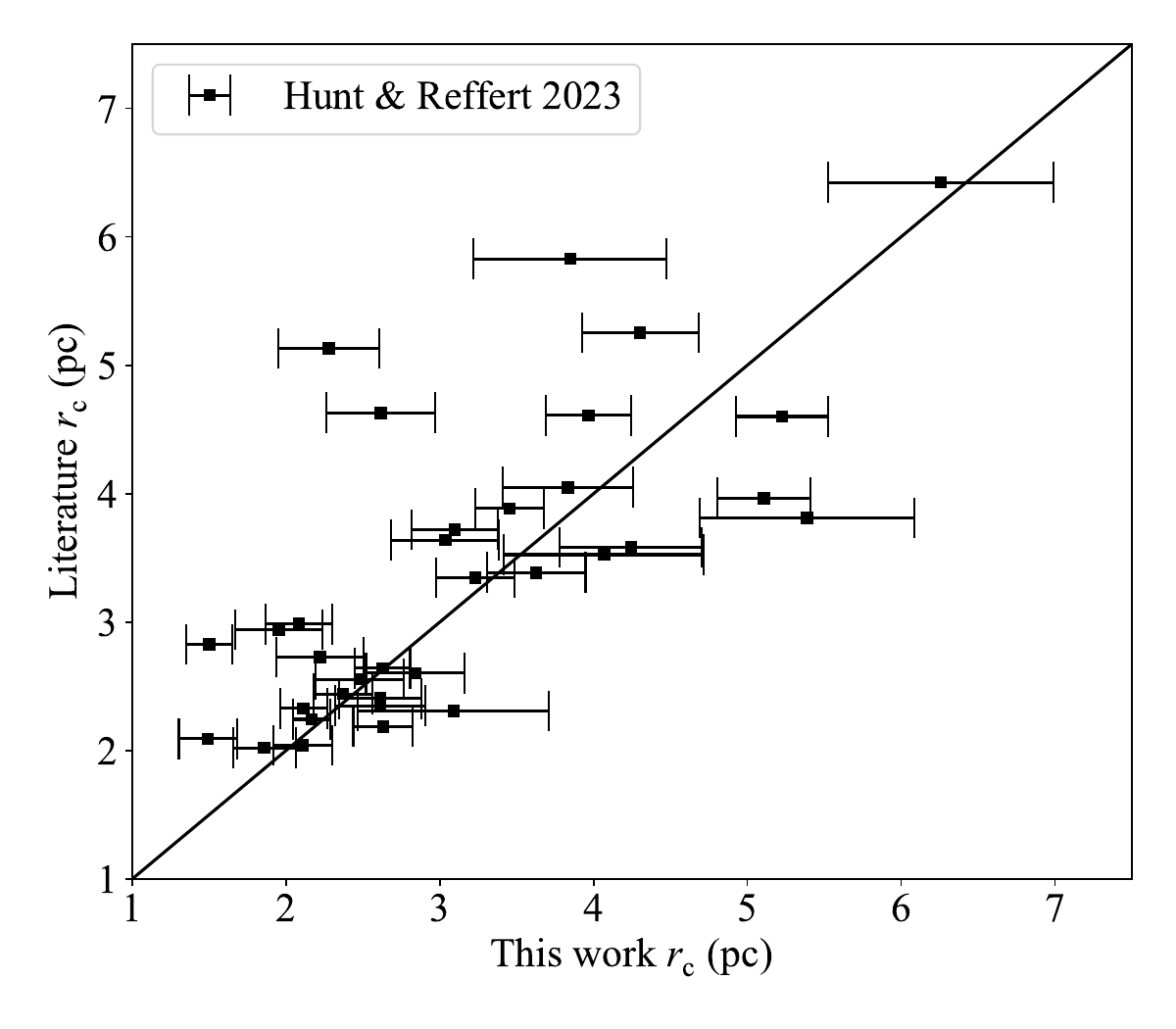}
\caption{A comparison of our derived cluster distances, reddening, ages and core radii to \cite{Dias2021} and \citep{Hunt2021}, where available. The 1:1 line is shown with a black diagonal.}
\label{fig:rc}
\end{figure*}

As an additional check on our BASE-9 derived cluster parameters, Figure \ref{fig:rc} compares our derived cluster parameters for distance, reddening and age with those from \cite{Hunt2023} and \cite{Dias2021}.  We find overall good agreement with the previous literature values.

 \begin{figure}
\includegraphics[width=.9\columnwidth]{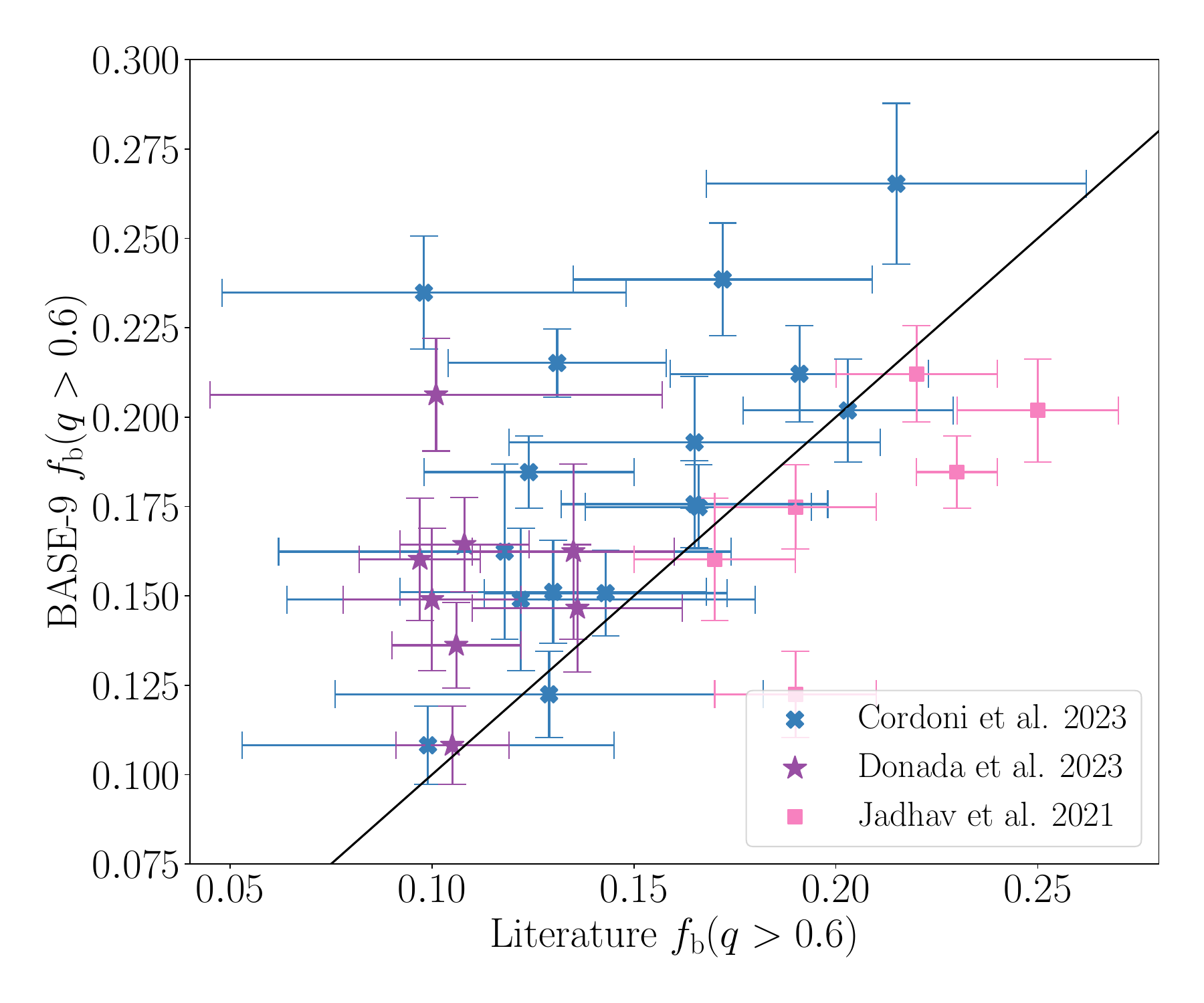}
\includegraphics[width=.9\columnwidth]{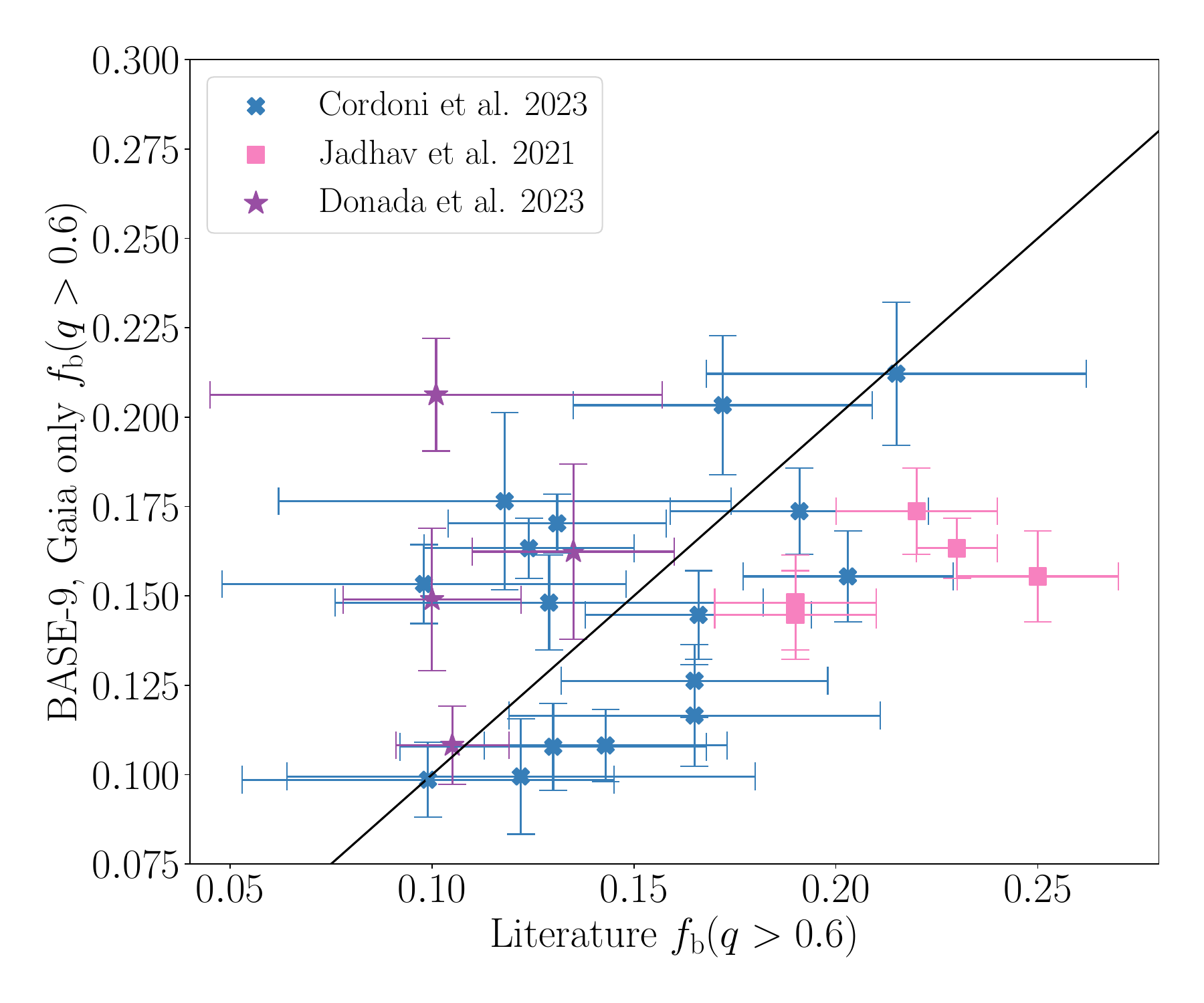}
\includegraphics[width=.9\columnwidth]{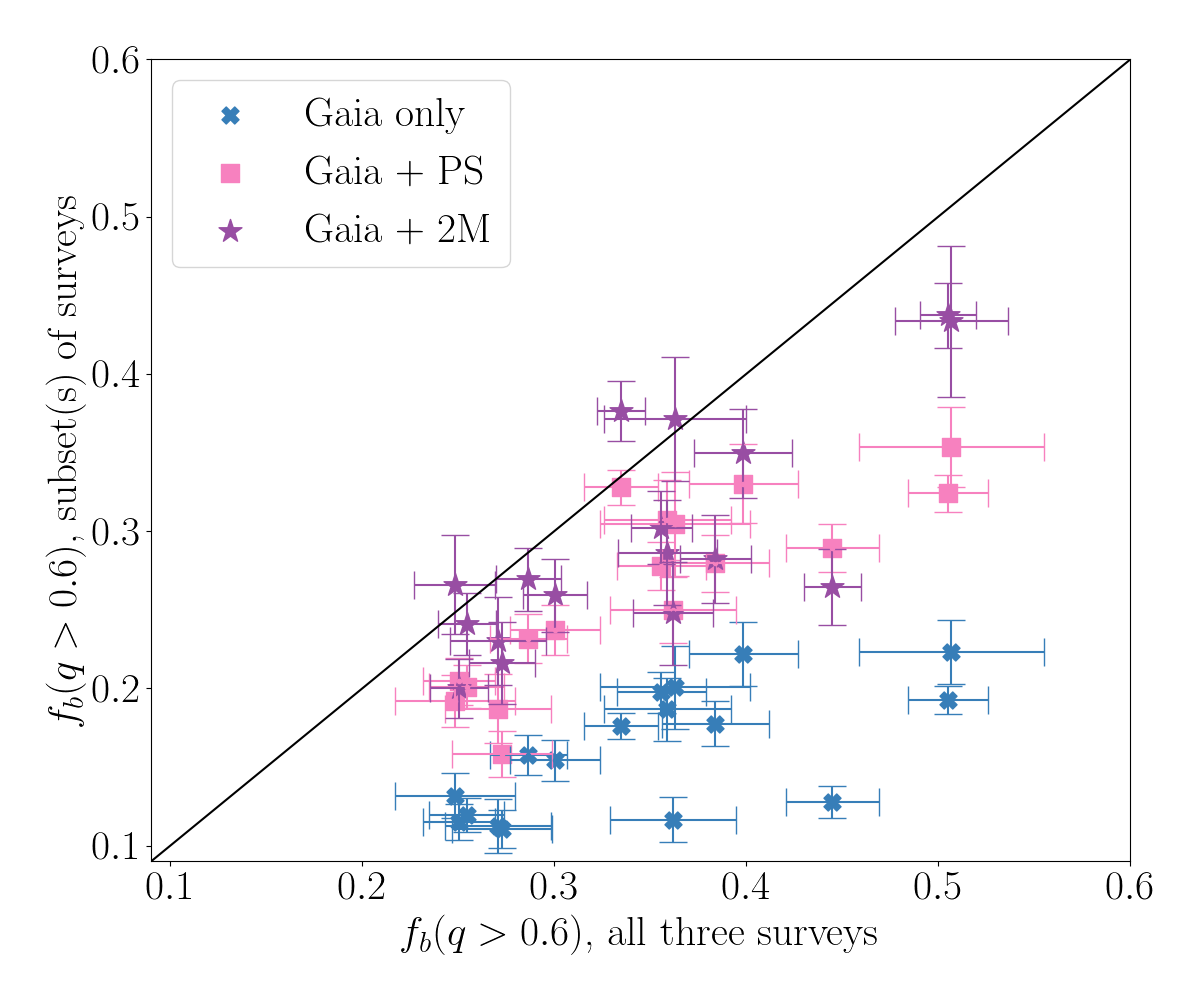}

    \caption{Top: Our binary fraction (BASE-9 $f_{\rm b}(q>0.6)$) with all 11 filters compared to the binary fractions from \cite{Jadhav2021, Cordoni2023, Donada2023} (Literature $f_{\rm b}(q>0.6)$). Middle: The same plot but our binary fractions are found using only Gaia photometry (BASE-9, Gaia only $f_{\rm b}(q>0.6)$). Bottom: The BASE-9 recovered binary fraction using photometry from all three surveys (horizontal axis) vs.\ the BASE-9 recovered binary fraction using only either Gaia, Gaia+PS, or Gaia+2M photometry (vertical axis). In all panels, the 1:1 line is shown with a black line. } 
    \label{fig:match}
\end{figure}

Table \ref{tab:cluster_params} also contains the spatial properties of the clusters.  We derive the cluster core and tidal radii, $r_{\rm c}$ and $r_{\rm t}$ respectively, using \cite{King1962} model fits for all cluster members within the effective radius ($r_{\rm eff}$) -- the maximum radius from the cluster center within which we consider stars for cluster membership.  We find that the fit parameters are sensitive to the effective radius, $r_{\rm eff}$, chosen for the cluster. Exterior to $r_{\rm eff}$ the likelihood of contamination from cluster non-members increases, and the cluster becomes indistinguishable from the field in our data.  To determine what value to use for $r_{\rm eff}$ we recursively increase $r_{\rm eff}$ in steps of $1^{\prime}$, starting from just outside the the cluster center, and for each $r_{\rm eff}$ value we apply a King model fit and calculate the $\chi^2$ value between the data and the King model fit.  We choose the $r_{\rm eff}$ value and corresponding King model fit where $\chi^2$ is minimized.  Figure \ref{fig:King_fits} shows the radial distribution of the cluster members out to $r_{\rm eff}$, the King model fit, and the resulting $r_{\rm c}$ and $r_{\rm t}$ for the clusters.  Table \ref{tab:cluster_params} lists $r_{\rm c}$, $r_{\rm t}$, and $r_{\rm eff}$ in arc minutes, and the half mass radius, $R_{\rm h}$, in degrees and parsecs.  Figure \ref{fig:rc} shows our $r_{\rm c}$ values versus those from \cite{Hunt2023}.  Although \cite{Hunt2023} does not provide uncertainties for their core radii approximations, we find good agreement with their values.

\begin{figure*}[hbt!]
\includegraphics[width=\columnwidth]{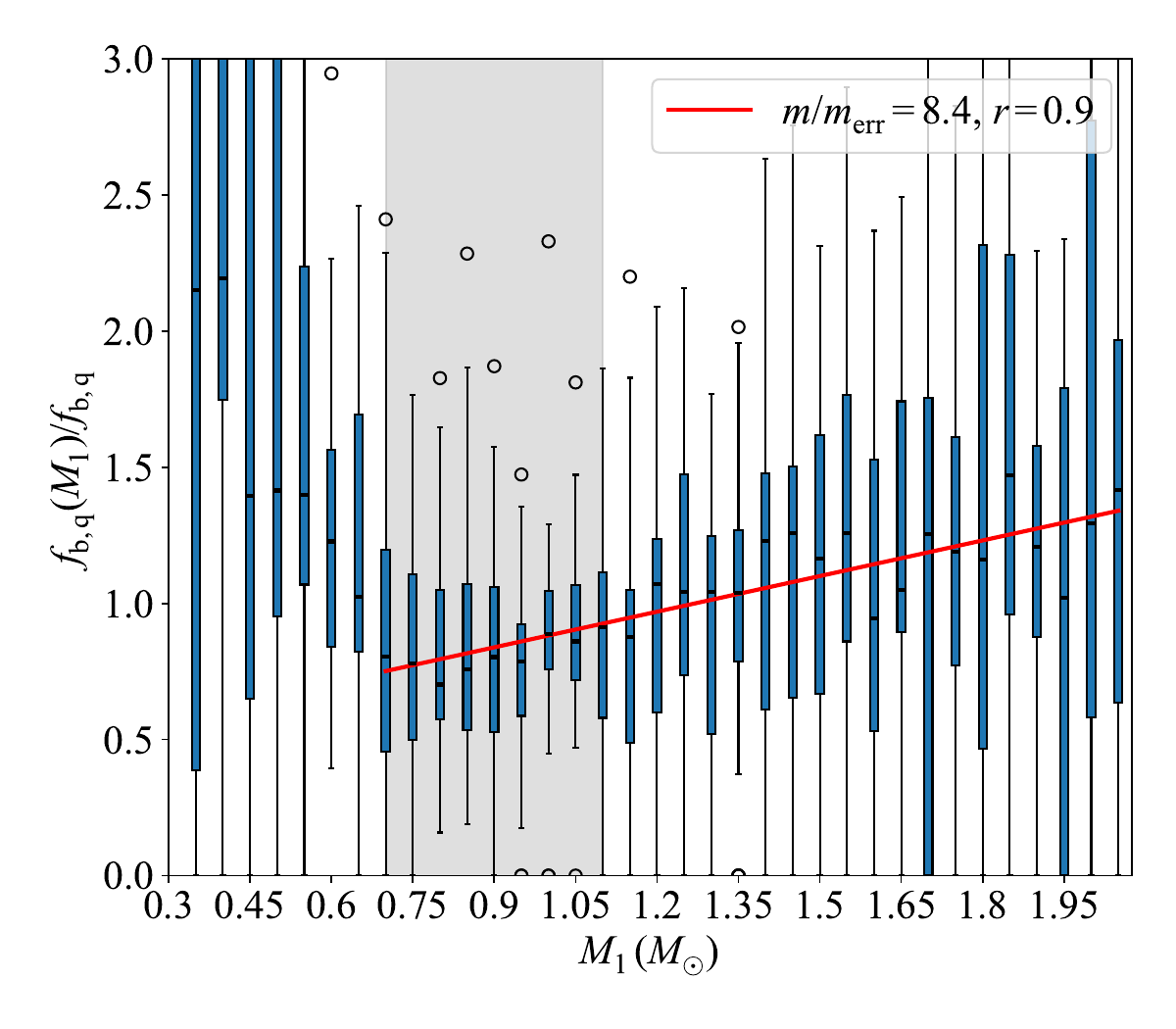}
\includegraphics[width=\columnwidth]{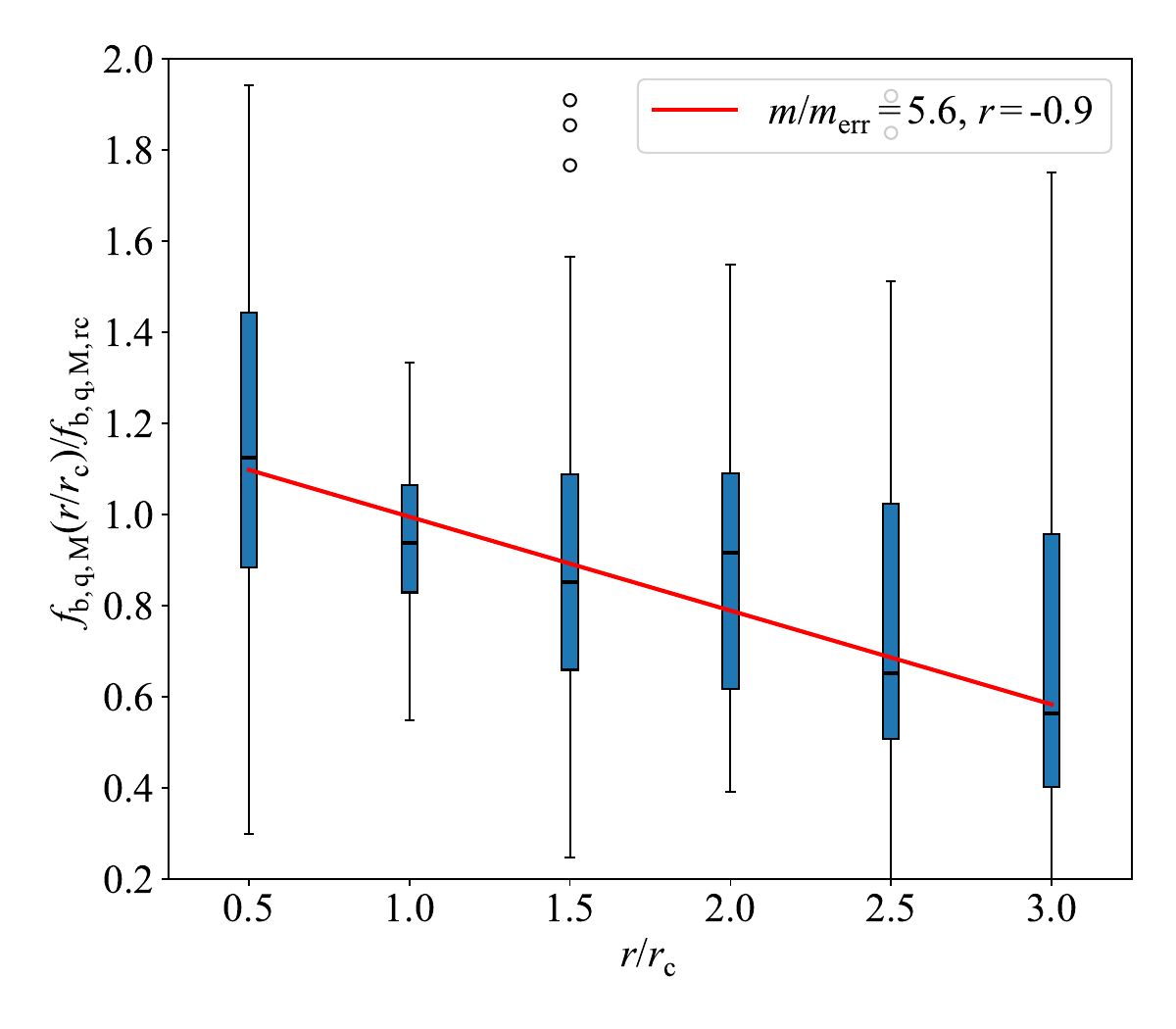}
    \caption{Left: Box and whisker plot of the binary fraction in bins of primary mass ($f_{\rm b,q}(M_1)$) normalized by the cluster's $f_{\rm b}$. Right: Box and whisker plot of \MLMSfb \, normalized by this binary fraction in the core \MLMSfbrc, in bins of distance from the cluster center (in units of core radius, $r_{\rm c}$). The line of best fit to the median of the bins is shown in red.} 
    \label{fig:fb_m1}
\end{figure*}

\subsection{Binary fractions and their dependence on photometric coverage}\label{sec:binary_fractions}
We compare our BASE-9 recovered photometric binary fractions to \citet{Jadhav2021}, \citet{Cordoni2023} and \citet{Donada2023} who have also looked at the photometric binary fraction in a number of OCs using only Gaia photometry.  For this comparison, we impose a limit of $q>0.6$ on our results to match the limit imposed by these authors.  The top panel of Figure \ref{fig:match} shows a subset of our BASE-9 derived binary fractions, for binaries with a mass ratio $q>0.6$, (BASE-9 $f_{\rm b}(q>0.6)$), versus the literature binary fractions with $q>0.6$ for the same clusters (labeled as Literature $f_{\rm b}(q>0.6)$ in the figure).    When we compare our values, we find that our binary fractions are consistently higher than the previous literature values (e.g., see the top panel of Figure~\ref{fig:match}).  We attribute this to a more successful recovery rate of the binaries due to the additional information from the Pan-STARRS1 (PS) and 2MASS (2M) photometry that we use.  

To test this, we run BASE-9 on this subset of clusters using only Gaia photometry (to match the photometric bands used in these comparison papers).  The middle panel of Figure \ref{fig:match} shows our binary fraction with only Gaia photometry (BASE-9 $f_{\rm b}(q>0.6),$ Gaia only) compared to the binary fractions from \citet{Jadhav2021}, \citet{Cordoni2023} and \citet{Donada2023}.  We show the 1:1 line with a black line.  Indeed, we find that using only Gaia photometry results in a lower binary fraction, and these BASE-9 binary fractions agree better with the literature values.  This result emphasizes the importance of using the largest wavelength coverage available when identifying photometric binaries as this can significantly increase the recovery rate of photometric binaries (sometimes up to 80\%).

To better understand if the increases in photometric binary recovery rate is coming from the five optical filters in PS or the three infrared filters in 2M, we determine the photometric binary fraction by giving BASE-9 Gaia+PS photometry or Gaia+2M photometry.  We plot the results in the bottom panel of Figure \ref{fig:match}.  As expected, each survey increases the recovery rate of binaries relative to the Gaia only sample, but we find that adding infrared photometry from 2M boosts the binary fraction more so than adding the optical PS photometry alone.  This may be because 2M observes in wavelengths longer than the Gaia red pass bands which provides useful additional information necessary to differentiate between binaries and singles (particularly for binaries with smaller mass ratios).  This supports the findings of \cite{Malofeeva2022, Malofeeva2023} who demonstrated that the binary track on a CMD becomes more distinct when using pseudo-colors comprised of optical and infrared bands rather than optical bands alone.

While 2MASS has a lower spatial resolution ($5^{"}$, \citealt{Skrutskie2006}) than Gaia ($1^{"}$, \citealt{Lindegren2021}), we have checked how blended 2MASS photometry from nearby Gaia sources might affect our results.  After searching each cluster member with 2MASS photometry for a nearby Gaia source, we find that no more than 1.5\% of an OC's members contain another Gaia source within a $2.5^{"}$ search radius.  We thus conclude that the differing spatial resolutions of the photometric surveys has a negligible affect on our results, and the increased binary recovery rate with 2MASS photometry is real.

\subsection{Defining a consistent binary sample}


Similar to Paper I, we aim to compare MS binary fractions over the same primary-mass range between clusters to look for correlations between binary fractions and cluster parameters. Again, in our implementation, BASE-9 will reject any object that cannot be fit by two normal non-degenerate stars along an isochrone (e.g., blue stragglers, sub-subgiants, white dwarfs) as non-members.  Though these objects are often found in binaries \citep{Mathieu2009, Mikhnevich2024}, we restrict our analysis here to a limited portion of the main-sequence (MS) track.  Note that some small portion of these MS binaries may be blue lurkers, which also appear to have a high binary fraction \citep{Leiner2019}, but these are not distinguishable from normal MS binaries using photometry alone.  

In order to self-consistently compare binary fractions between the different OCs, we impose a set of restrictions on the binary sample we consider. We only consider cluster members within three core radii, binaries with a mass ratio ($q$) greater than 0.5, and define a mass-limited -- MS (ML-MS) sample that must contain at least 100 members ($N_{\rm ML-MS}$).  We describe how we arrived at each cut in turn here.  We refer to the binary fraction for cluster members within this range as \MLMSfb, and the binary fraction for these stars that reside within the cluster's core radius as \MLMSfbrc.  Our goal is to then compare these binary fractions between clusters of varying characteristics to search for trends.



\subsubsection{Binary fraction vs. primary mass} 
\cite{Cohen2020} and Paper I found that BASE-9 does a much better job at accurately recovering binaries with $q> 0.5$ since binaries with lower $q$ are easy to confuse with single stars.  Thus, for each cluster we report the binary fraction for only the binaries with $q >0.5$ so that we may compare $f_{\rm b}$ consistently between many different clusters. We refer to the binary fraction that only considers binaries with $q>0.5$ as \fb.

In the left panel in Figure \ref{fig:fb_m1} we show a box and whisker plot of the binary fraction for a given primary mass, $f_{\rm b}(M_1)$; the number of binaries within the $M_1$ bin divided by the number of cluster members within the same bin, normalized by the cluster's \fb \, (within three core radii for reasons described in the following section).  For primary masses $\geq0.7$, we observe an increasing binary fraction with increasing primary mass.  This result is similar to Paper I (which included a subset of these clusters, see Section \ref{subsection:cluster_sample} for details).  Observations of binaries in the field also show that the binary fraction increases with primary mass \citep{Raghavan2010, Duchene2013}.  We show the line of best fit to the median values of the bins with $M_1 > 0.7 \, M_{\odot}$ in red.  We find a strong correlation ($r=0.9$, $m/m_{\rm err}=8.4$) between primary mass and binary fraction.  For $M_1 \lesssim 0.7$ we observe the opposite trend (binary fraction increasing toward lower masses).  We attribute this primarily to field star contamination as the smallest stars have the dimmest and least certain photometry and kinematic measurements. We note that some observations suggest that the mass-ratio distribution is shifted to higher mass ratios as one moves toward lower primary masses \citep{Goodwin2013}; however there is some disagreement in the literature on this point; e.g. \citet{Duchene2013} conclude that the mass-ratio distribution is flat for all primary masses $>0.3M_\odot$.  If the $q$ distribution truly is shifted toward higher values for lower primary mass, this increase in binary fraction that we observe for $M_1 < 0.7 M_\odot$ may have a physical basis, because binaries with higher $q$ are more stable against disruption.  Nonetheless, as the increase we observe could be due to field star contamination, we set our minimum primary mass cut to $0.7 \, M_{\odot}$, where $f_{\rm b,q}(M_1)/f_{\rm b,q}$ is near the lowest in the left panel of Figure~\ref{fig:fb_m1}, in order to limit field-star contamination.  To choose our upper mass limit we find the maximum mass that is below the turn off for most of the OCs in our sample.  We find this mass to be $1.1 \, M_{\odot}$.  

The mass range for our ML-MS sample is thus $0.7 - 1.1 \, M_{\odot}$, and we mark this range in Figure \ref{fig:fb_m1} with the gray shaded region. We refer to the binary fraction within in the ML-MS range for binaries with $q>0.5$ as \MLMSfb.  We mark this ML-MS bound for each cluster in Figure \ref{fig:CMDs} with black horizontal lines. The exception to this mass range is the two oldest OCs in our sample, Berkeley 39 and NGC 188, whose turnoff happens at lower masses.  Thus, for Berkeley 39 and NGC 188 we impose upper mass limits of $1 \, M_{\odot}$ and $0.98 \, M_{\odot}$, respectively.  The upper mass limits for these two OCs are shown with red dashed lines in Figure \ref{fig:CMDs}.   For the majority of our subsequent analysis, we only consider the clusters for which there are at least 100 $N_{\rm ML-MS}$ in order to ensure a more reliable binary fraction.  31/35 of the OCs meet this criteria and we refer to this subset of our full sample as the primary sample.  

\subsubsection{Binary fraction vs. distance from cluster center}

In the right panel of Figure \ref{fig:fb_m1} we show, for each OC, \MLMSfb \, (the total binary fraction with $q>0.5$ in the ML-MS range) normalized by the ML-MS binary fraction $q>0.5$ in the cluster core (\MLMSfbrc) in bins of distance away from the cluster center (in units of $r_{\rm c}$).  We plot a line of best fit to the median (normalized) binary fractions in each bin and find a very strong anti-correlation ($r=-0.9$ and $m/m_{\rm err}=5.6)$ between binary fraction and distance from the cluster center.  The concentration of binaries towards the cluster center is also observed in globular clusters \citep{Milone2012} and a similar result was found in Paper I for OCs, although for a more limited sample.  We discuss this result in the context of binary formation,  dynamical evolution and mass segregation in Section \ref{sec:Discussion}.

We expect field star contamination to increase with distance from the OC center, given that the OC stellar density falls off with distance while the field stellar density remains roughly constant (over the region of sky covered by a given OC).  Furthermore, because the binaries occupy a larger area on a CMD than single stars \citep{Spangler2025RNAASS}, we expect field-star contamination to most adversely affect our binary population, thus raising the binary fraction. Indeed, in some clusters we observe that our binary fraction increases when analyzing stars well beyond 3 core radii from the cluster center.  In order to limit this affect we only consider stars with $r < 3r_c$ in the OCs when calculating \MLMSfb\ in our subsequent analysis. 

\subsection{Binary fraction and cluster parameter correlations}
In Figure \ref{fig:allOC_map} we show a correlation matrix for \MLMSfb\ and other cluster parameters for the OCs in our primary sample.  If we take the Pearson's correlation coefficient $|r| \geq 0.5$ to indicate a strong correlation, we find a significant correlation between \MLMSfb\ (and \MLMSfbrc) only with the dynamical age of the cluster, Age ($t_{\rm rh}$).

 \begin{figure*}[hbt!]
\includegraphics[width=1\columnwidth]{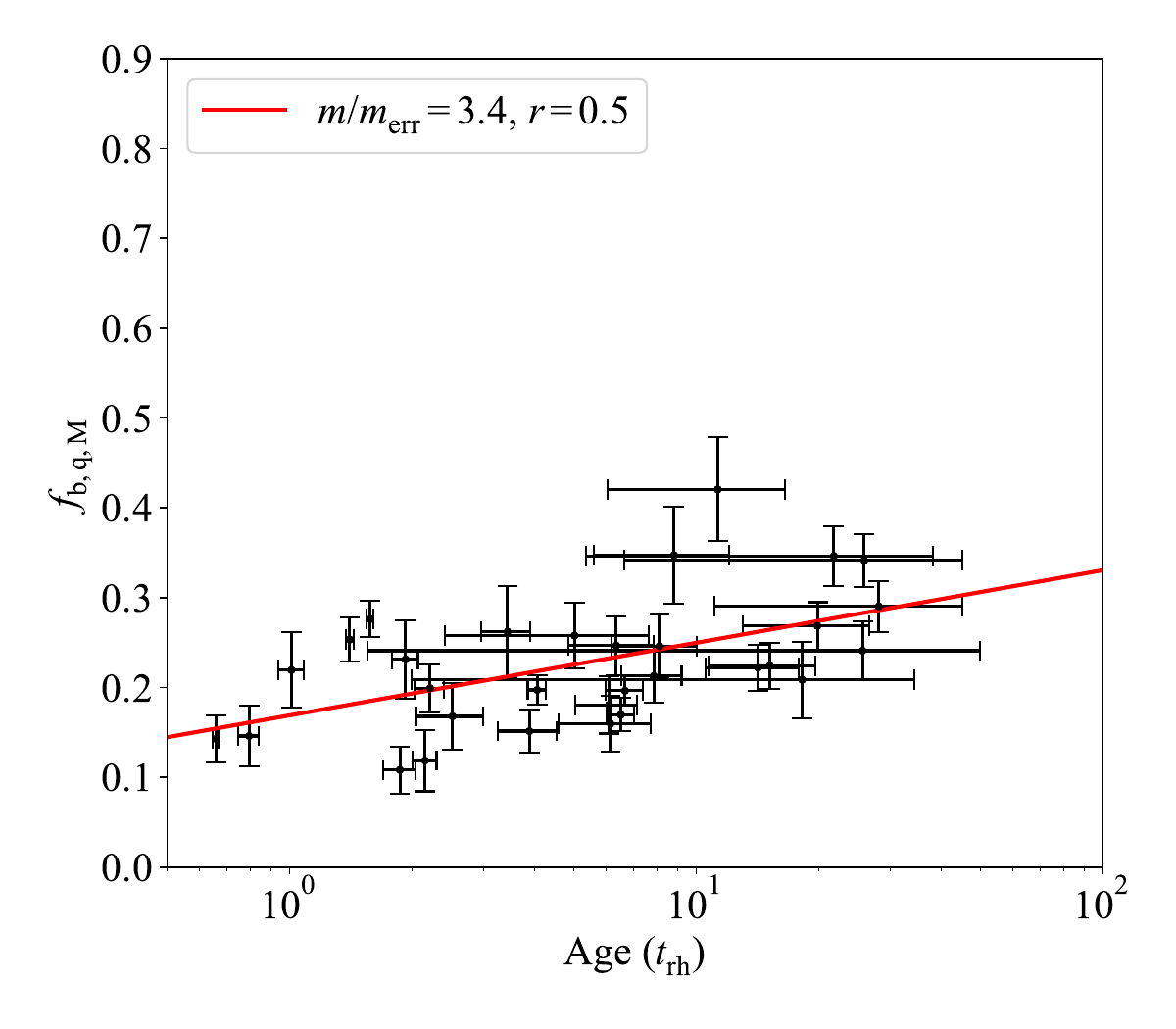}
\includegraphics[width=1\columnwidth]{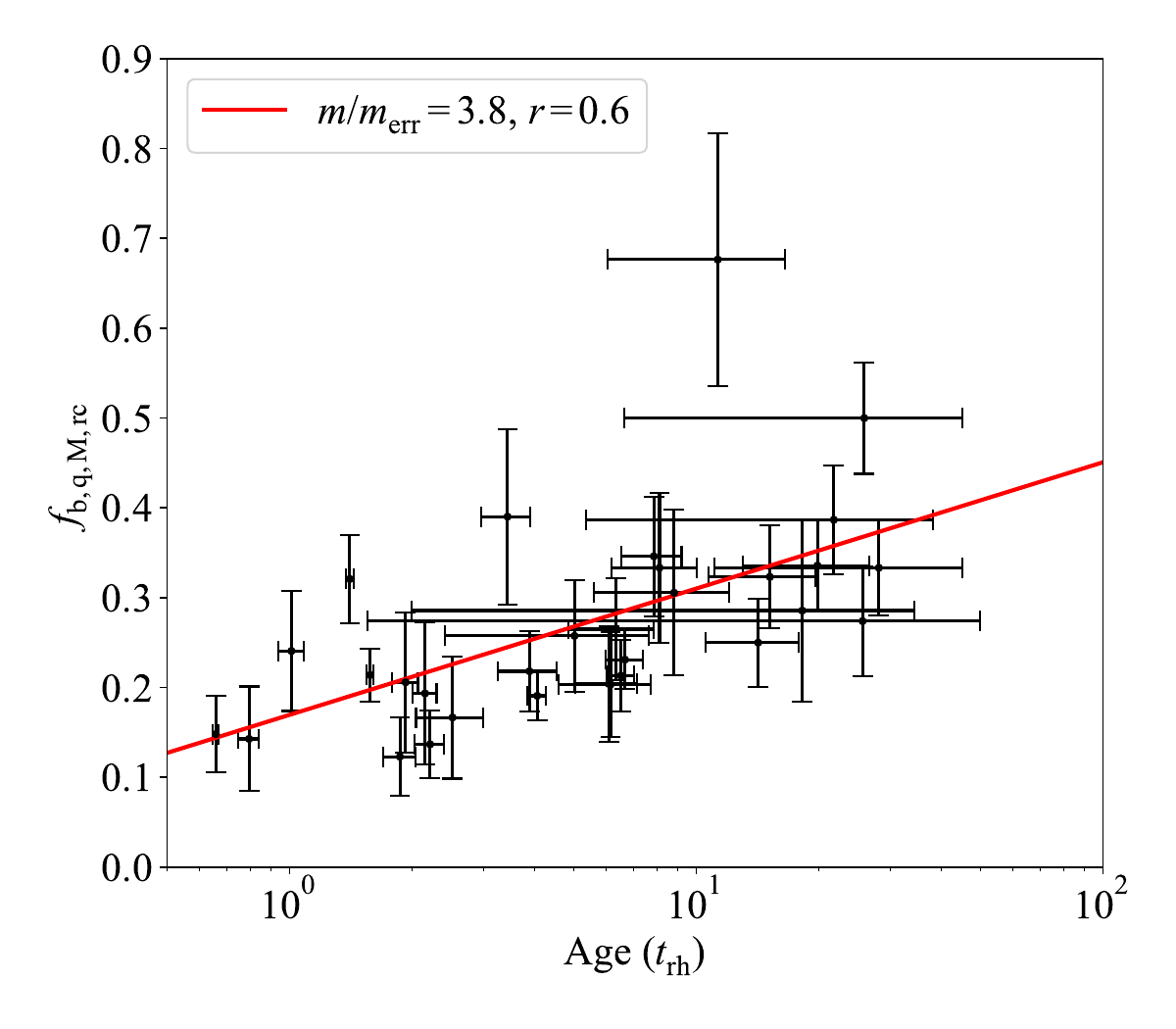}

    \caption{Left: \MLMSfb\ versus OC dynamical age for our primary sample. Right: \MLMSfbrc\ versus OC dynamical age for our primary sample.  The line of best fit is shown in red.} 
    \label{fig:fb_v_age}
\end{figure*}


 

The dynamical age is the number of half-mass relaxation timescales the cluster has lived through.  The half-mass relaxation timescale is
\begin{equation}
    t_{\rm rh}=\frac{0.346 N (r_{\rm c,3D})^{3/2}}{\sqrt{GM_{\rm Tot}}ln \Lambda},
\end{equation}
where $N$ is the total number of objects (single or higher order star systems) in the cluster, $G$ is the gravitational constant, $M_{\rm Tot}$ is the total mass of the cluster and $\Lambda$ is Coulombs constant which we assume to be $0.1 N$ \citep{Giersz1994,Heggie2003}.  

The left panel of Figure \ref{fig:fb_v_age} shows OC dynamical age versus \MLMSfb\ for our primary sample with the line of best fit in red.  The linear fit has a Pearson's correlation coefficient of $r=0.5$, with a $t$-statistic of $m/m_{\rm err}=3.4$.  The trend is even stronger when considering only the binary fraction in the core of the OC.  The right panel of Figure \ref{fig:fb_v_age} shows OC dynamical age versus \MLMSfbrc.  The linear fit has a Pearson's correlation coefficient of $r=0.6$, with a $t-$statistic of $m/m_{\rm err}=3.8$ (the correlation is slightly stronger than shown in the correlation matrix of Figure \ref{fig:allOC_map} when considering age on a log scale).

\subsection{Binary mass-ratio distributions}
\begin{figure*}[hbt!]

\includegraphics[width=.33\textwidth]{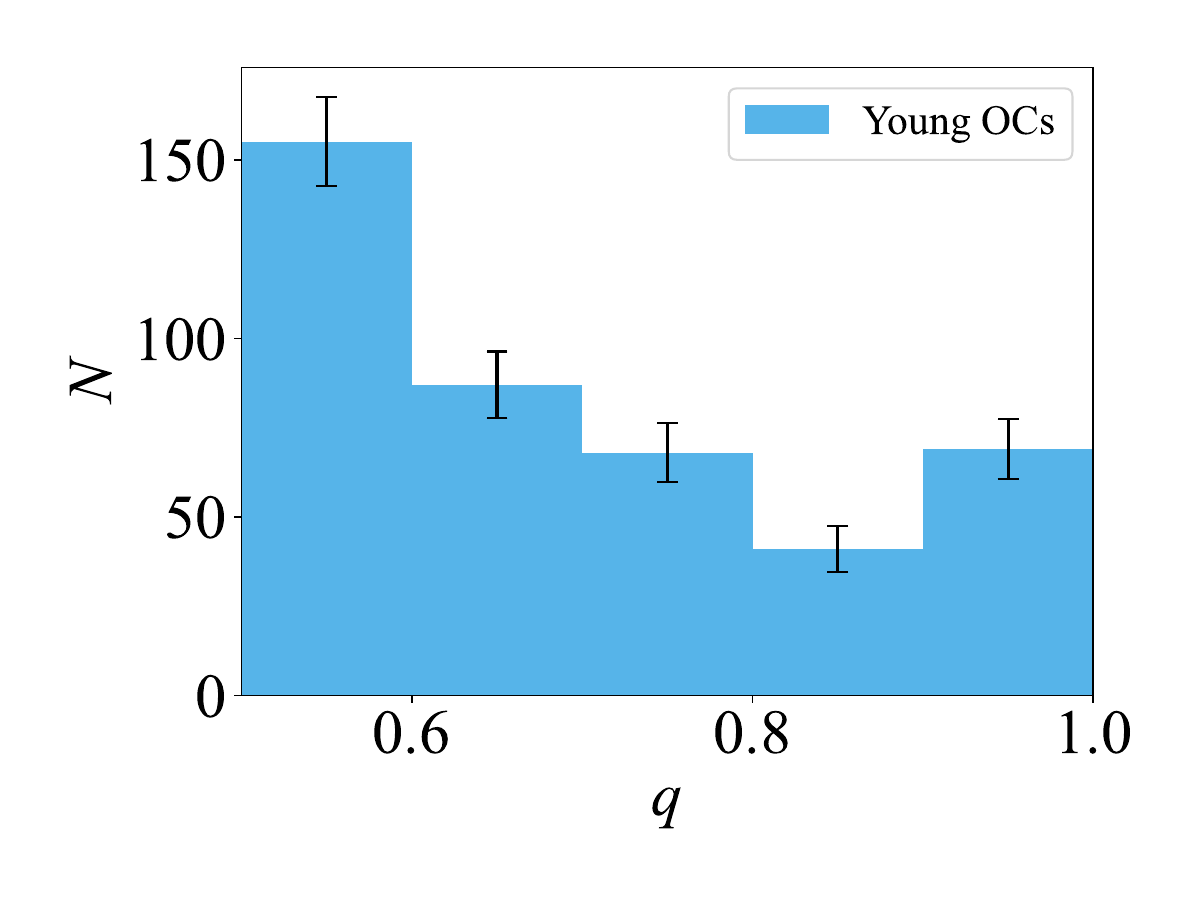}
\includegraphics[width=.33\textwidth]{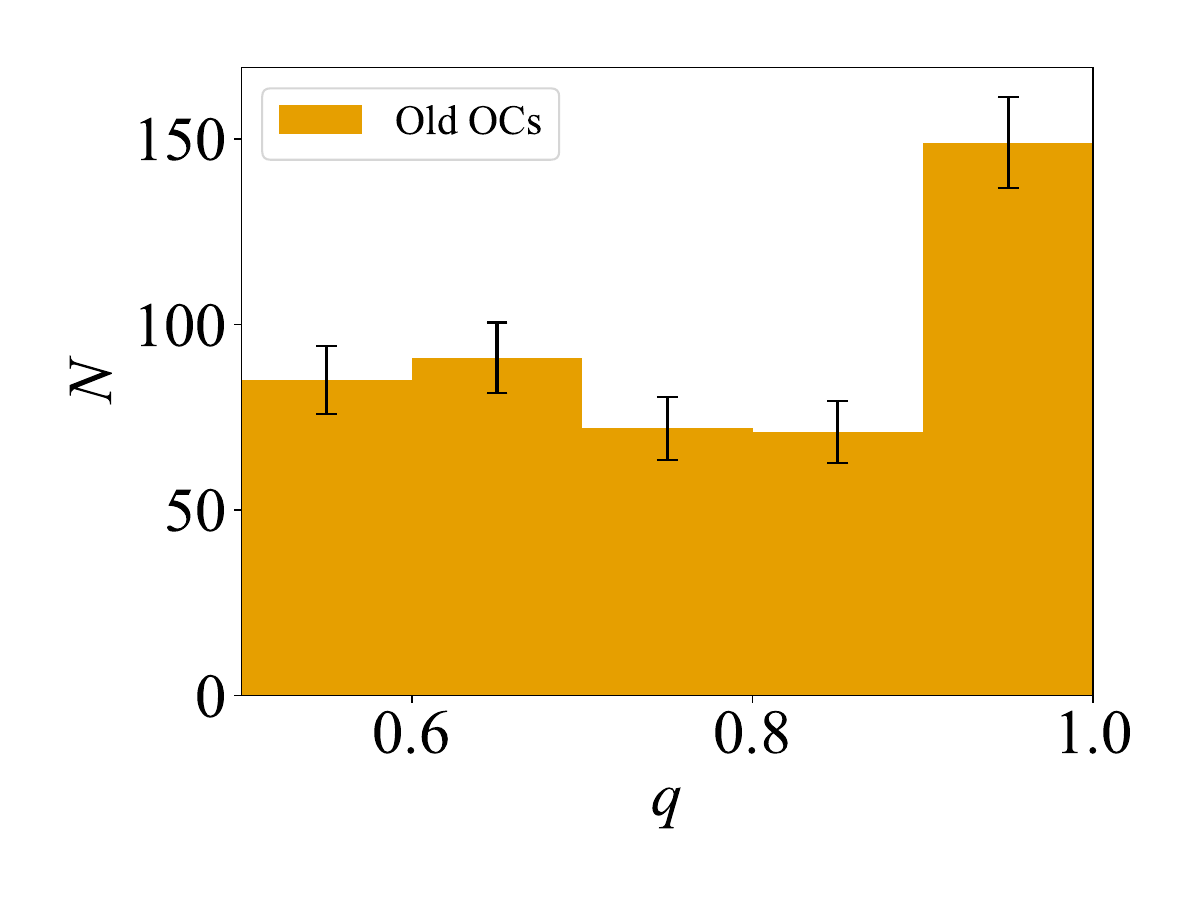}
\includegraphics[width=.33\textwidth]{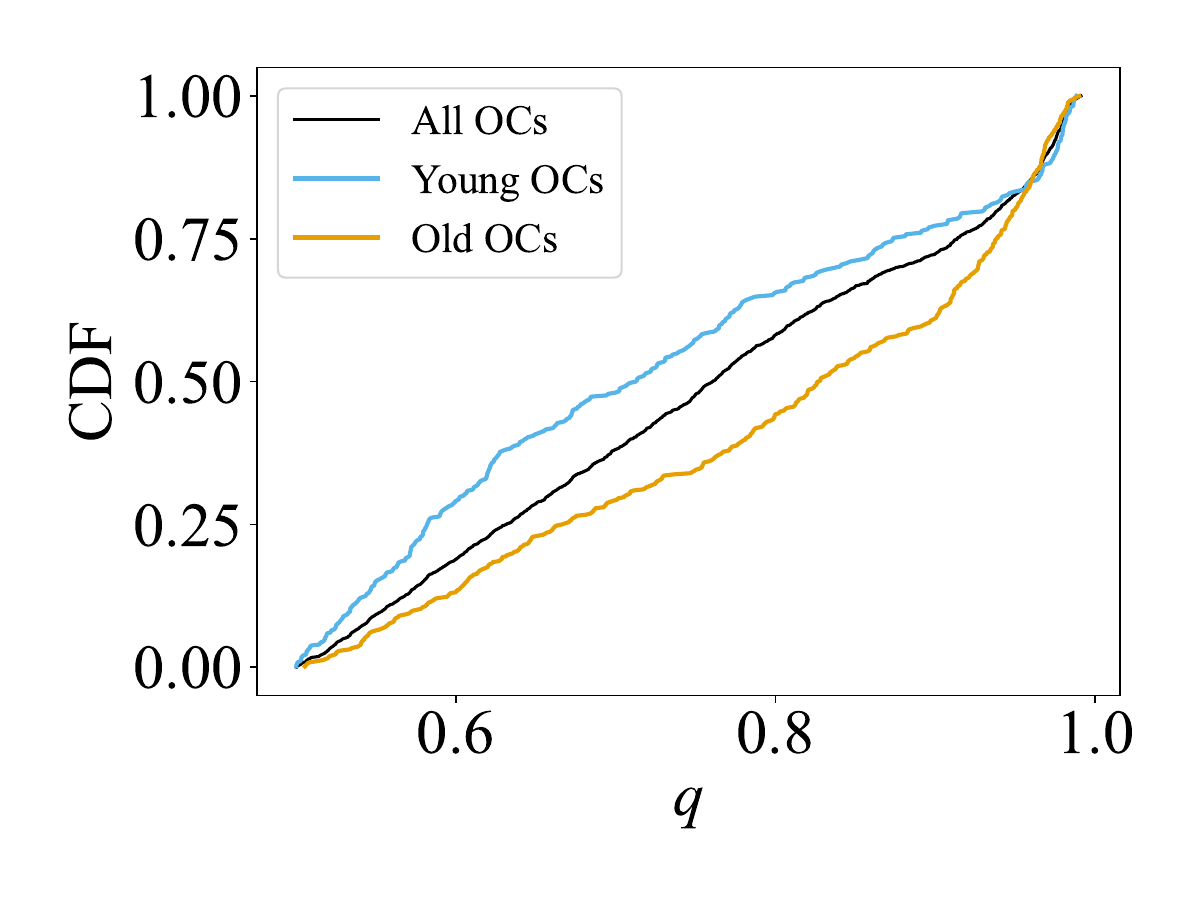}

    \caption{We find statistically distinct distributions of $q>0.5$ between dynamically young (blue) and dynamically old (yellow) OCs.  Left panel: $q$ distribution for ML-MS high mass ratio binaries in seven dynamically young clusters with ages less than $2 \, t_{\rm rh}$. Middle panel: $q$ distribution for ML-MS high mass ratio binaries in nine dynamically old clusters with ages greater than $15 \, t_{\rm rh}$. Right panel: CDFs of the $q$-distributions for all 31 ML-MS high mass ratio binaries within three core radii (black), as compared to the distributions for the youngest (blue) and oldest (yellow) OCs.}
    \label{fig:cluster_qs}
\end{figure*}


Figure \ref{fig:all_q_tr} in the Appendix shows the ML-MS $q$ distribution within three core radii for all \noclusters\, OCs in our full sample in order of increasing dynamical age.  We find that in general, dynamically young clusters show more variation in their $q$ distributions while dynamically old clusters have a distribution more consistent with uniform, except for a peak near $q=1$. As an illustrative example, we show the combined ML-MS $q$-distribution within three core radii for seven of the dynamically youngest clusters with ages less than $2 \, t_{\rm rh}$ and for seven of the dynamically oldest clusters that have ages greater than $15 \, t_{\rm rh}$ in our primary sample in the left and middle panels of Figure \ref{fig:cluster_qs}, respectively.  The older OC's have a flat distribution with a peak near $q=1$.  This peak at $q\sim1$ is also observed in the field for solar-type binaries \citep{Raghavan2010}. The younger OCs display a ``valley" distribution where $q$ decreases from 0.5 until about $q\approx0.9$ where it then begins to increase.  The right panel of Figure \ref{fig:cluster_qs} shows these same data as cumulative distribution functions (CDFs) for an easier direct comparison.  The mass-ratio distributions between dynamically young and old OCs are statistically distinct with a Kolmogorov–Smirnov (K-S) $p$-value of $\sim 9\times10^{-12}$ and an Anderson-Darling $p$-value of 0.001.

To try to quantify this ``valley" distribution we perform a Hartigan Dip Test on the ML-MS $q>0.5$ distributions in our primary sample.  The Hartigan Dip Test of Unimodality is a statistical test for hypothesizing the modality of a distribution and returns a Dip statistic and corresponding $p$-value \citep{Hartigan_1985}.  The Dip statistic is the distance of the data's CDF away from a favorable unimodal CDF, and the $p$-value assesses the statistical significance of the Dip statistic.  
An OC with a mass-ratio distribution that cannot be distinguished from uniform should have a low Dip statistic and a high $p$-value, whereas an OC that has the ``valley" distribution would be classified as multimodal and thus have a large Dip statistic and low $p$-value.

\begin{figure*}[hbt!]
\includegraphics[width=1\columnwidth]{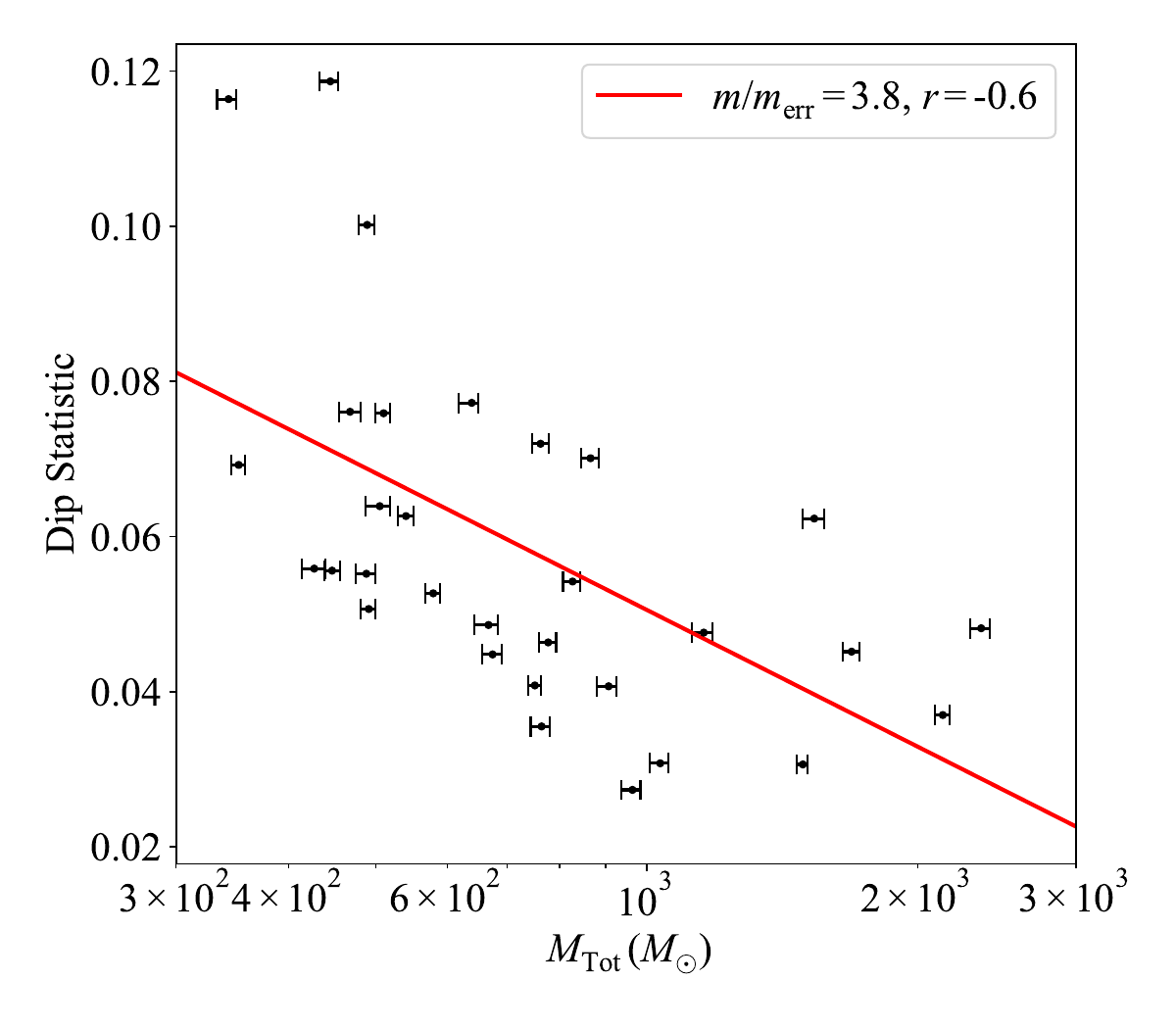}
\includegraphics[width=1\columnwidth]{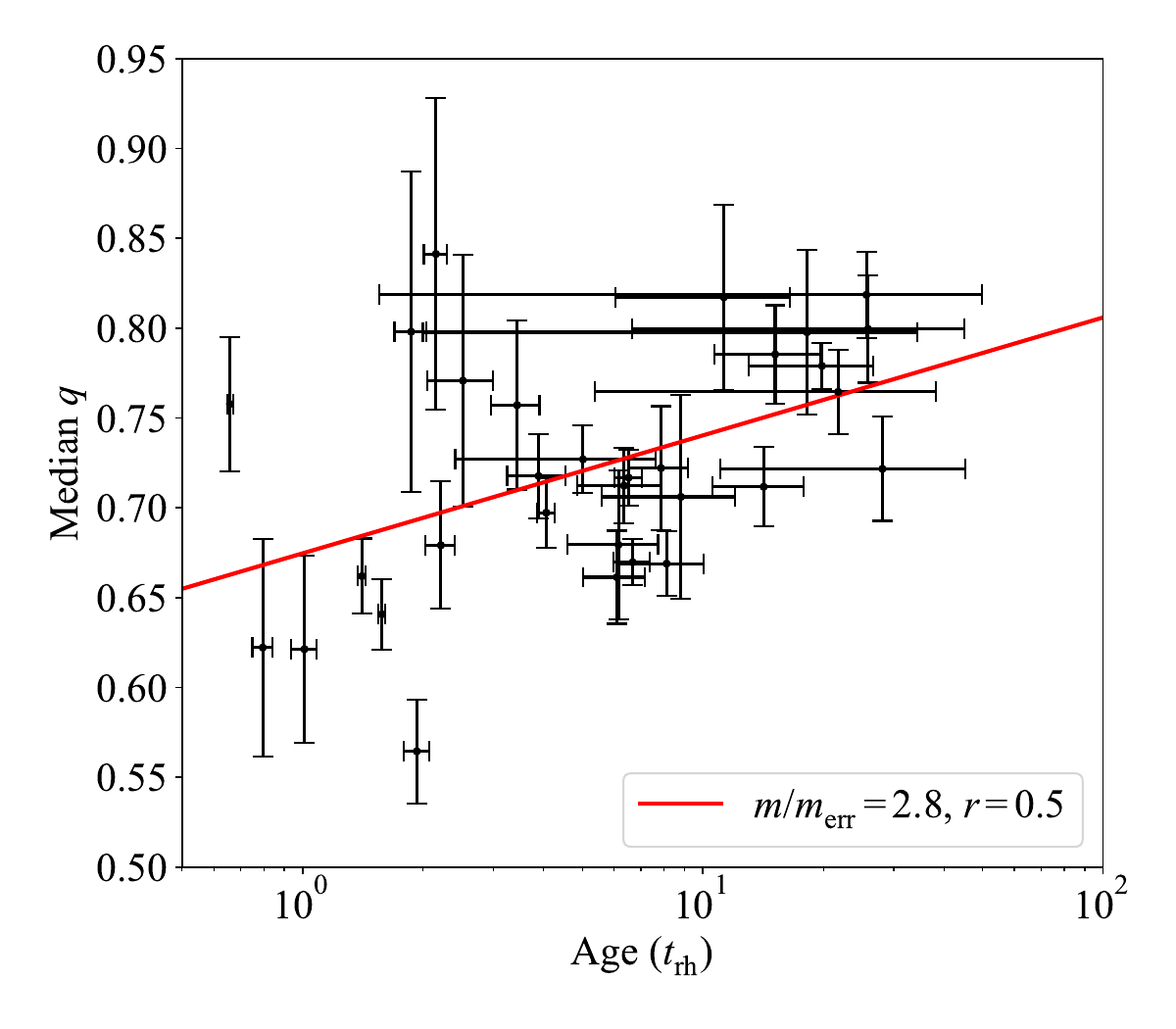}

    \caption{Left: Total OC mass versus the Dip statistic for the $q$ distribution in our primary sample. Right: OC dynamical age versus median $q$ value in our primary sample.} 
    \label{fig:dip_values}
\end{figure*}

After calculating the Dip statistic for the primary sample, we look for correlations between the Dip statistic and various cluster parameters.  We find a strong anti-correlations between the Dip statistic and the total cluster mass ($r=-0.6$ and $m/m_{\rm err}=3.8$), shown in the left panel of Figure \ref{fig:dip_values}, indicating that the binary mass-ratio distribution becomes more unimodal as total cluster mass increases. 

The right panel of Figure \ref{fig:dip_values} shows a strong correlation with the median $q$ of the \MLMSfb binaries and cluster dynamical age ($r=0.5$ and $m/m_{\rm err}=2.8$). When considering the median $q$-value within only the core (\MLMSfbrc), we find an even stronger correlation with $r=0.6$ and $m/m_{\rm err}=3.8$.   We provide our interpretation of these results and discuss their implications in the following section.

\section{Discussion}\label{sec:Discussion}

As we summarized in Section~\ref{sec:intro}, stellar dynamical and relaxation processes are expected to modify the binary population over time.  In this section we connect our observations to these theoretical predictions and offer our interpretation of these results. 

Theoretical predictions suggest that the MS binary fraction should increase in the cluster core relative to the cluster halo over time due to mass segregation \citep{Cote1991, Layden_1999, Fregeau_2002}.  The left panel of Figure \ref{fig:fb_m1} shows an increasing binary fraction toward the core, when considering data from all OCs, as would be expected from mass segregation.  This finding is consistent with previous work that took a closer look at a subset of the OCs considered here \citep{Childs2024, Motherway2024, Zwicker2024}.  Another expected consequence of mass segregation when combined with cluster evaporation processes is the preferential loss of low mass stars, which would tend to increase the overall binary fraction of a cluster over time.  The left panel of Figure \ref{fig:fb_v_age} shows a strong correlation between \MLMSfb\ and dynamical age for the 31 OCs and an even stronger correlation when considering only the binary fraction in the core, as seen in the right panel of Figure \ref{fig:fb_v_age}.  We take all these findings together to be strong evidence of mass segregation operating in our sample of OCs.

We also find a correlation between the shape of the binary mass-ratio distribution and the total cluster mass and cluster dynamical age.  To quantify this, we performed the Hartigan Dip-test \citep{Hartigan_1985} which is a measure of likelihood that the data are drawn from a unimodal distribution. We find that OCs with lower total mass have a larger Dip Statistic for their  mass-ratio distributions, indicating that they tend toward a multi-modal distribution.  On the other hand, more massive OCs have a smaller Dip Statistic, indicating that these mass-ratio distributions tend towards a unimodal distribution (e.g., see Figure \ref{fig:cluster_qs}).
Furthermore, in the right panel of Figure \ref{fig:dip_values} we find a strong positive correlation between dynamical age and the median mass ratio.  This result is even stronger when only considering binaries in the core (where the encounter rate is expected to be highest); in this sample we find the median $q$ to correlate with the dynamical age of the OC with Pearson's correlation coefficient of $r=0.6$ and $m/m_{\rm err}=3.8$.  In our data, more dynamically evolved binary populations tend toward higher mass-ratio values and a more unimodal (often uniform with a peak near $q=1$) mass-ratio distribution.

We interpret the above findings to be evidence of dynamical modification of the binary mass ratios over time, and specifically the preferential loss of low-mass-ratio binaries.  Encounters are expected to be more frequent and more energetic within more massive clusters and those with higher velocity dispersions. As discussed in Section~\ref{sec:intro}, close encounters in dense stellar environments are predicted to preferentially disrupt wide and low-mass-ratio binaries, establishing a “hard-soft” boundary \citep{heggie75, Geller_2013}.  Additionally, stellar encounters can modify the mass-ratio distribution of a binary population as exchanges tend to pair together the most massive stars in an encounter \citep{heggie75}.  Our findings indicate that over time, these exchange encounters tend to shift the binary mass-ratio distribution to higher mass ratios.  (While a particular encounter between a low-mass high-q binary and a more massive star can result in a binary with a lower $q$, future encounters of this binary with similar mass stars would likely result in a higher $q$ binary.)  All of these processes may contribute to explaining our finding, as a combination of the preferential dynamical destruction of wide and low-mass-ratio binaries and exchange encounters that tend to increase a binary's mass ratio. These effects become more pronounced in dynamically older and more massive OCs.  

Interestingly, this empirical result is also consistent with observations of similar binaries in the field.  For solar-type binaries in the field, the mass-ratio distribution is observed to be uniform except for an apparent excess of ``twins" ($q$=1) \citep{Raghavan2010}.   This may suggest that many solar-type field binaries originated in OCs and have gone through dynamical modifications, which favor large $q$ binaries, before being dispersed to the field.

We have also investigated the distribution of the system mass of the binaries (i.e., primary plus secondary mass) as a function of dynamical age.  \citet{Goodwin2013} suggests that if the binary population is primordial, the system mass is a more fundamental parameter to probe, than for instance the mass ratio, as the system mass can yield information on the binary formation process.   In a primordial binary population, the system mass distribution originating from molecular cloud fragmentation is expected to be log-normal ($\mu=-0.7, \sigma=0.6)$ over $0.2-6 \, M_{\odot}$ \citep{Chabrier2003}.   We might then expect the system mass distribution to increasingly deviate from a log-normal distribution with dynamical age if dynamics are indeed changing the primordial companions.  We find that none of our OCs have a system-mass distribution that is consistent with the \citet{Chabrier2003} log-normal distribution (limited to be within our ML-MS mass range).  Additionally, we do not find a trend between the binary system-mass distribution and OC dynamical age in our data.  We note that our data are incomplete for systems with $q<0.5$, and the \citet{Chabrier2003} log-normal distribution extends far beyond the mass range we consider in this paper, so further investigation into this question with additional empirical datasets is warranted.

Finally, we emphasize that because our sample size is relatively small, the dynamical interpretations we discuss here should be taken with caution.  Additionally, BASE-9 currently does not have the ability to accurately model higher ordered systems (with more than two stars).  If there is a triple star system and the combined light of the stars places the system between the usual single and equal-mass-binary sequences on the MS track, BASE-9 may find a single or binary star solution that fits the object and, if so, it may be included in our member sample.  As a result, there may be undetected/unknown tertiaries for a subset of the binaries in our sample.  Given that tertiaries are prominent in both the field \citep{Tokovinin2014} and in OCs \citep{Malofeeva2023}, our photometric binaries may provide interesting samples for follow-up observations searching for triples in OCs.  Lastly, more theoretical work is desirable to more fully understand the dynamical processes that drive the formation and dynamical evolution of the $q$ distribution of OCs; our observational results will be an important touchstone for future work of this kind.

\section{Summary \& Conclusions}\label{sec:Conclusions}
We  use the BASE-9 code with Gaia DR3, Pan-STARRS, and 2MASS photometry to constrain cluster parameters, binarity and individual stellar masses of the cluster members for \noclusters\,OCs.  The OCs in this sample reside between $500-4000\, \rm pc$ and have $A_{\rm V}<1$.  Our derived cluster metallicities are consistent with APOGEE measurements and our derived ages are consistent with the known $|Z_{\rm GC}|$ correlation.  We are able to recover more binaries than previous studies due to our inclusion of more photometry, specifically in the infrared.  

Considering our full OC sample as a whole, we find that binaries are preferentially found towards the cluster center and the binary fraction increases with primary mass. We search for trends between cluster parameters and binary fractions of main-sequence stars that have a primary mass of $0.7-1.1 \, M_{\odot}$, as this is where our data is most complete between all OCs in our sample.  In our sample the binary fraction increases with OC dynamical age, and the correlation is even stronger when considering only the binary fraction in the core -- an expected outcome of mass segregation.   

We find the binary mass-ratio ($q$) distribution of dynamically young OCs is statistically distinct from that of the old OCs.  On average, dynamically young OCs display multi-modal $q$ distributions rising toward unity and toward our detection limit of $q=0.5$ while more dynamically evolved clusters display more uniform $q$ distributions often with a peak near $q=1$.   Interestingly, the uniform $q$ distribution with a peak near $q=1$ is consistent with binaries in the field.  We also observe a similar transition from multi-modal to unimodal $q$ distributions when comparing low mass to high mass OCs in our sample.  

Lastly, we find a correlation between the median $q$ of the binary population in a cluster and the cluster dynamical age.   We attribute this to dynamical processing, and specifically to a combination of stellar exchanges (commonly resulting in partners of more similar masses) and the destruction of softer binaries that may have a lower mass ratios than hard binaries that are preserved.  The similar $q$-distributions found in dynamically old OCs and the field suggests that many solar-type field binaries originated in OCs and have gone through dynamical modifications, which favor large $q$ binaries, before being dispersed to the field.

In summary, we find evidence for both mass segregation effects and strong dynamical encounters shaping the binary population in these OCs.  Importantly, here we leverage precise constraints on the binary mass-ratios, which are readily available from this and other photometric binary studies (and much harder to derive from other methods of binary detections).  However, our sample size is relatively small, and thus our conclusions should be approached with caution.  Larger samples would be helpful to confirm our findings (though may be difficult to construct given larger distances, higher reddening, etc.).  The data used in this paper are available on Zenodo \citep{Childs_Zenodo} and can also be explored interactively in our OCbinaryexplorer website at \href{https://ocbinaryexplorer.ciera.northwestern.edu/}{https://ocbinaryexplorer.ciera.northwestern.edu/}.

\begin{acknowledgments}
We thank the anonymous referee for comments that have improved this manuscript.  We thank Ted von Hippel, Elliot Robinson, Elizabeth Jeffery, Emily Leiner, David Stenning, and Bill Jefferys for helpful conversations that have improved this manuscript.  ACC and AMG acknowledge support from the National Science Foundation (NSF) under grant No. AST-2107738. Any opinions, findings, and conclusions or recommendations expressed in this material are those of the author(s) and do not necessarily reflect the views of the NSF. This research was supported in part through the computational resources and staff contributions provided for the Quest high performance computing facility at Northwestern University which is jointly supported by the Office of the Provost, the Office for Research, and Northwestern University Information Technology.  

\end{acknowledgments}

%

\vspace{5mm}
\facilities{Gaia DR3 (DOI:10.5270). The Pan-
STARRS1 Surveys (PS1) and the PS1 public science
archive have been made possible through contributions
by the Institute for Astronomy, the University of Hawaii,
the Pan-STARRS Project Office, the Max-Planck Soci-
ety and its participating institutes, the Max Planck In-
stitute for Astronomy, Heidelberg and the Max Planck
Institute for Extraterrestrial Physics, Garching, The
Johns Hopkins University, Durham University, the Uni-
versity of Edinburgh, the Queen’s University Belfast,
the Harvard-Smithsonian Center for Astrophysics, the
Las Cumbres Observatory Global Telescope Network
Incorporated, the National Central University of Tai-
wan, the Space Telescope Science Institute, the National
Aeronautics and Space Administration under Grant No.
NNX08AR22G issued through the Planetary Science Di-
vision of the NASA Science Mission Directorate, the
National Science Foundation Grant No. AST-1238877,
the University of Maryland, Eotvos Lorand University
(ELTE), the Los Alamos National Laboratory, and the
Gordon and Betty Moore Foundation. This publication
makes use of data products from the Two Micron All
Sky Survey, which is a joint project of the University of
Massachusetts and the Infrared Processing and Analysis Center/California Institute of Technology, funded by
the National Aeronautics and Space Administration and
the National Science Foundation.  Funding for the Sloan Digital Sky Survey IV has been provided by the Alfred P. Sloan Foundation, the U.S. Department of Energy Office of Science, and the Participating Institutions. SDSS acknowledges support and resources from the Center for High-Performance Computing at the University of Utah. The SDSS website is www.sdss.org.}

\software{
Astropy \citep{astropy:2013, astropy:2018, astropy:2022}}

\appendix

\section{Additional plots and tables}

\begin{turnpage}
\begin{table}
\caption{OC parameters.}
\label{tab:cluster_params}
\resizebox{\linewidth}{!}{%
\begin{tabular}{lcccccccccccccccccc}
\toprule
Cluster & min $p$-value & Age/Myr & Distance/pc & [Fe/H] & $A_{\rm V}$/mmag & $R_{\rm h}^{\circ}$ & $R_{\rm h}$/pc & $r_{\rm eff}^{\prime}$ & $r_{\rm c}^{\prime}$ & $r_{\rm t}^{\prime}$ & $M/M_{\odot}$ & $N$ & $N_{\rm c}$ & $t_{\rm rh}$/Myr & Age/$t_{\rm rh}$ & $f_{\rm{b,q,M}}$ & $f_{\rm{b,q,M,rc}}$ & Median $q$ \\
\midrule

Trumpler 3 & 0.002 & $50^{+1}_{-0}$ & $677^{+324}_{--313}$ & $0.16^{+0.01}_{-0.01}$ & $836^{+9}_{-6}$ & 0.14 & 1.63 & 150 & 5.7 & 394.2 & $252^{+5}_{-5}$ & 525 & 57 & $78.0^{+19.0}_{-17.0}$ & $0.6^{+0.0}_{-0.0}$ & $0.31^{+0.08}_{-0.08}$ & $0.31^{+0.14}_{-0.14}$ & $0.71^{+0.08}_{-0.08}$ \\
Collinder 394 & 0.288 & $159^{+1}_{-1}$ & $600^{+265}_{--258}$ & $-0.44^{+0.01}_{-0.01}$ & $786^{+4}_{-4}$ & 0.34 & 3.55 & 70 & 16.5 & 143.5 & $468^{+13}_{-13}$ & 416 & 123 & $200.0^{+64.0}_{-54.0}$ & $0.8^{+0.0}_{-0.0}$ & $0.15^{+0.03}_{-0.03}$ & $0.14^{+0.06}_{-0.06}$ & $0.62^{+0.06}_{-0.06}$ \\
NGC 2301 & 0.315 & $184^{+2}_{-2}$ & $793^{+116}_{--110}$ & $-0.24^{+0.01}_{-0.01}$ & $289^{+3}_{-3}$ & 0.12 & 1.59 & 100 & 4.9 & 197.5 & $343^{+7}_{-10}$ & 711 & 92 & $85.0^{+17.0}_{-15.0}$ & $2.2^{+0.1}_{-0.1}$ & $0.12^{+0.03}_{-0.03}$ & $0.19^{+0.08}_{-0.08}$ & $0.84^{+0.09}_{-0.09}$ \\
NGC 2323 & 0.448 & $190^{+0}_{-1}$ & $1008^{+377}_{--369}$ & $0.13^{+0.01}_{-0.01}$ & $683^{+4}_{-4}$ & 0.12 & 2.1 & 73 & 5.3 & 121.3 & $505^{+14}_{-18}$ & 578 & 90 & $98.0^{+21.0}_{-19.0}$ & $1.9^{+0.1}_{-0.1}$ & $0.23^{+0.04}_{-0.04}$ & $0.21^{+0.08}_{-0.08}$ & $0.56^{+0.03}_{-0.03}$ \\
NGC 2287 & 0.002 & $190^{+0}_{-0}$ & $751^{+60}_{--58}$ & $0.08^{+0.00}_{-0.00}$ & $164^{+2}_{-2}$ & 0.27 & 3.57 & 70 & 14.8 & 83.5 & $866^{+19}_{-21}$ & 806 & 283 & $287.0^{+52.0}_{-47.0}$ & $0.7^{+0.0}_{-0.0}$ & $0.14^{+0.03}_{-0.03}$ & $0.15^{+0.04}_{-0.04}$ & $0.76^{+0.04}_{-0.04}$ \\
NGC 7243 & 0.002 & $193^{+1}_{-0}$ & $909^{+325}_{--318}$ & $0.14^{+0.01}_{-0.01}$ & $658^{+4}_{-4}$ & 0.19 & 3.02 & 40 & 11.5 & 55.5 & $488^{+11}_{-13}$ & 357 & 144 & $191.0^{+60.0}_{-51.0}$ & $1.0^{+0.1}_{-0.1}$ & $0.22^{+0.04}_{-0.04}$ & $0.24^{+0.07}_{-0.07}$ & $0.62^{+0.05}_{-0.05}$ \\
NGC 2168 & 0.059 & $336^{+3}_{-5}$ & $871^{+306}_{--302}$ & $0.06^{+0.00}_{-0.00}$ & $651^{+2}_{-3}$ & 0.21 & 3.2 & 70 & 10.2 & 100.3 & $1535^{+38}_{-44}$ & 1401 & 366 & $239.0^{+33.0}_{-30.0}$ & $1.4^{+0.0}_{-0.0}$ & $0.25^{+0.02}_{-0.02}$ & $0.32^{+0.05}_{-0.05}$ & $0.66^{+0.02}_{-0.02}$ \\
NGC 1912 & 0.092 & $526^{+7}_{-3}$ & $1157^{+570}_{--554}$ & $0.09^{+0.01}_{-0.00}$ & $858^{+5}_{-5}$ & 0.19 & 3.74 & 37 & 9.5 & 62.6 & $1035^{+22}_{-27}$ & 725 & 255 & $237.0^{+51.0}_{-46.0}$ & $2.2^{+0.2}_{-0.2}$ & $0.20^{+0.03}_{-0.03}$ & $0.14^{+0.04}_{-0.04}$ & $0.68^{+0.04}_{-0.04}$ \\
NGC 1664 & 0.002 & $577^{+9}_{-4}$ & $1402^{+582}_{--565}$ & $0.25^{+0.01}_{-0.01}$ & $745^{+6}_{-6}$ & 0.1 & 2.52 & 15 & 5.7 & 30.0 & $579^{+10}_{-11}$ & 353 & 156 & $115.0^{+44.0}_{-37.0}$ & $5.0^{+2.6}_{-2.6}$ & $0.26^{+0.04}_{-0.04}$ & $0.26^{+0.06}_{-0.06}$ & $0.73^{+0.02}_{-0.02}$ \\
NGC 2281 & 0.002 & $602^{+2}_{-1}$ & $545^{+74}_{--72}$ & $0.13^{+0.00}_{-0.00}$ & $274^{+2}_{-3}$ & 0.28 & 2.63 & 80 & 13.4 & 125.2 & $447^{+9}_{-9}$ & 620 & 157 & $175.0^{+39.0}_{-35.0}$ & $3.4^{+0.5}_{-0.5}$ & $0.26^{+0.05}_{-0.05}$ & $0.39^{+0.10}_{-0.10}$ & $0.76^{+0.05}_{-0.05}$ \\
NGC 7209 & 0.002 & $674^{+3}_{-3}$ & $1100^{+313}_{--304}$ & $-0.10^{+0.01}_{-0.01}$ & $537^{+4}_{-4}$ & 0.23 & 4.38 & 50 & 11.4 & 86.4 & $445^{+9}_{-12}$ & 370 & 112 & $268.0^{+86.0}_{-73.0}$ & $2.5^{+0.5}_{-0.5}$ & $0.17^{+0.04}_{-0.04}$ & $0.17^{+0.07}_{-0.07}$ & $0.77^{+0.07}_{-0.07}$ \\
NGC 2548 & 0.043 & $675^{+1}_{-1}$ & $759^{+56}_{--54}$ & $-0.01^{+0.00}_{-0.01}$ & $152^{+1}_{-2}$ & 0.33 & 4.43 & 80 & 18.3 & 104.7 & $489^{+9}_{-10}$ & 480 & 176 & $361.0^{+92.0}_{-80.0}$ & $1.9^{+0.2}_{-0.2}$ & $0.11^{+0.03}_{-0.03}$ & $0.12^{+0.04}_{-0.04}$ & $0.80^{+0.09}_{-0.09}$ \\
NGC 2437 & 0.083 & $723^{+7}_{-6}$ & $1514^{+401}_{--381}$ & $-0.13^{+0.01}_{-0.01}$ & $498^{+4}_{-4}$ & 0.18 & 4.77 & 50 & 8.9 & 68.6 & $2133^{+37}_{-43}$ & 1782 & 562 & $458.0^{+55.0}_{-50.0}$ & $1.6^{+0.0}_{-0.0}$ & $0.28^{+0.02}_{-0.02}$ & $0.21^{+0.03}_{-0.03}$ & $0.64^{+0.02}_{-0.02}$ \\
NGC 2447 & 0.002 & $757^{+0}_{-1}$ & $938^{+63}_{--61}$ & $-0.29^{+0.00}_{-0.00}$ & $140^{+2}_{-3}$ & 0.12 & 1.95 & 42 & 5.8 & 81.7 & $674^{+16}_{-18}$ & 924 & 212 & $124.0^{+23.0}_{-21.0}$ & $6.1^{+1.1}_{-1.1}$ & $0.18^{+0.03}_{-0.03}$ & $0.20^{+0.06}_{-0.06}$ & $0.66^{+0.03}_{-0.03}$ \\
NGC 2355 & 0.002 & $953^{+1}_{-1}$ & $2084^{+442}_{--434}$ & $0.18^{+0.00}_{-0.00}$ & $415^{+3}_{-2}$ & 0.06 & 2.1 & 30 & 3.1 & 44.7 & $510^{+9}_{-11}$ & 470 & 112 & $108.0^{+26.0}_{-22.0}$ & $8.8^{+3.2}_{-3.2}$ & $0.35^{+0.05}_{-0.05}$ & $0.31^{+0.09}_{-0.09}$ & $0.71^{+0.06}_{-0.06}$ \\
NGC 2099 & 0.014 & $959^{+1}_{-0}$ & $1254^{+551}_{--549}$ & $-0.22^{+0.00}_{-0.00}$ & $790^{+2}_{-2}$ & 0.1 & 2.11 & 50 & 4.5 & 81.3 & $1689^{+33}_{-39}$ & 1946 & 384 & $147.0^{+17.0}_{-16.0}$ & $6.5^{+0.5}_{-0.5}$ & $0.17^{+0.02}_{-0.02}$ & $0.21^{+0.04}_{-0.04}$ & $0.72^{+0.02}_{-0.02}$ \\
NGC 2539 & 0.045 & $967^{+1}_{-2}$ & $1190^{+99}_{--94}$ & $-0.12^{+0.01}_{-0.01}$ & $169^{+4}_{-4}$ & 0.14 & 2.81 & 40 & 6.6 & 67.1 & $540^{+10}_{-11}$ & 535 & 144 & $157.0^{+36.0}_{-32.0}$ & $6.2^{+1.6}_{-1.6}$ & $0.16^{+0.03}_{-0.03}$ & $0.20^{+0.06}_{-0.06}$ & $0.68^{+0.04}_{-0.04}$ \\
NGC 2423 & 0.006 & $1069^{+0}_{-0}$ & $943^{+182}_{--180}$ & $0.01^{+0.00}_{-0.00}$ & $381^{+3}_{-4}$ & 0.13 & 2.22 & 60 & 6.2 & 140.0 & $354^{+9}_{-11}$ & 418 & 79 & $105.0^{+30.0}_{-26.0}$ & $10.2^{+6.3}_{-6.3}$ & $0.27^{+0.05}_{-0.05}$ & $0.23^{+0.09}_{-0.09}$ & $0.79^{+0.04}_{-0.04}$ \\
NGC 2360 & 0.002 & $1069^{+1}_{-0}$ & $1114^{+250}_{--245}$ & $-0.00^{+0.00}_{-0.00}$ & $436^{+2}_{-2}$ & 0.11 & 2.1 & 50 & 5.5 & 77.5 & $827^{+17}_{-20}$ & 951 & 214 & $136.0^{+23.0}_{-20.0}$ & $7.9^{+1.3}_{-1.3}$ & $0.21^{+0.03}_{-0.03}$ & $0.35^{+0.07}_{-0.07}$ & $0.72^{+0.03}_{-0.03}$ \\
NGC 1245 & 0.007 & $1331^{+6}_{-5}$ & $2680^{+957}_{--936}$ & $0.01^{+0.01}_{-0.01}$ & $657^{+5}_{-5}$ & 0.06 & 2.67 & 25 & 3.0 & 33.3 & $764^{+17}_{-21}$ & 681 & 182 & $164.0^{+31.0}_{-28.0}$ & $8.1^{+1.9}_{-1.9}$ & $0.25^{+0.04}_{-0.04}$ & $0.33^{+0.08}_{-0.08}$ & $0.67^{+0.02}_{-0.02}$ \\
NGC 6940 & 0.002 & $1347^{+2}_{-2}$ & $955^{+301}_{--296}$ & $-0.11^{+0.00}_{-0.00}$ & $591^{+4}_{-4}$ & 0.18 & 3.03 & 40 & 10.4 & 56.9 & $762^{+16}_{-17}$ & 627 & 245 & $212.0^{+46.0}_{-41.0}$ & $6.4^{+1.5}_{-1.5}$ & $0.25^{+0.03}_{-0.03}$ & $0.27^{+0.06}_{-0.06}$ & $0.71^{+0.02}_{-0.02}$ \\
Tombaugh 1 & 0.042 & $1502^{+2}_{-2}$ & $2370^{+1007}_{--996}$ & $-0.24^{+0.00}_{-0.00}$ & $765^{+4}_{-3}$ & 0.06 & 2.28 & 40 & 2.8 & 50.1 & $427^{+12}_{-13}$ & 491 & 99 & $133.0^{+30.0}_{-27.0}$ & $11.3^{+5.2}_{-5.2}$ & $0.42^{+0.06}_{-0.06}$ & $0.68^{+0.14}_{-0.14}$ & $0.82^{+0.05}_{-0.05}$ \\
NGC 1817 & 0.046 & $1719^{+73}_{-7}$ & $1401^{+538}_{--508}$ & $-0.46^{+0.01}_{-0.06}$ & $699^{+10}_{-5}$ & 0.19 & 4.7 & 40 & 11.5 & 44.1 & $639^{+11}_{-21}$ & 556 & 251 & $442.0^{+105.0}_{-90.0}$ & $3.9^{+0.6}_{-0.6}$ & $0.15^{+0.02}_{-0.02}$ & $0.22^{+0.04}_{-0.04}$ & $0.72^{+0.02}_{-0.02}$ \\
NGC 2627 & 0.163 & $1893^{+3}_{-1}$ & $1842^{+347}_{--336}$ & $-0.07^{+0.00}_{-0.00}$ & $369^{+2}_{-4}$ & 0.06 & 2.03 & 30 & 3.1 & 56.2 & $352^{+6}_{-7}$ & 431 & 84 & $104.0^{+26.0}_{-23.0}$ & $18.2^{+16.2}_{-16.2}$ & $0.21^{+0.04}_{-0.04}$ & $0.29^{+0.10}_{-0.10}$ & $0.80^{+0.05}_{-0.05}$ \\
Melotte 71 & 0.004 & $1901^{+1}_{-0}$ & $1781^{+537}_{--526}$ & $-0.62^{+0.01}_{-0.01}$ & $569^{+3}_{-5}$ & 0.07 & 2.04 & 41 & 3.1 & 53.5 & $751^{+11}_{-13}$ & 1091 & 209 & $134.0^{+20.0}_{-18.0}$ & $14.2^{+3.6}_{-3.6}$ & $0.22^{+0.03}_{-0.03}$ & $0.25^{+0.05}_{-0.05}$ & $0.71^{+0.02}_{-0.02}$ \\
NGC 7789 & 0.274 & $1901^{+0}_{-0}$ & $1859^{+937}_{--930}$ & $-0.05^{+0.00}_{-0.00}$ & $884^{+2}_{-2}$ & 0.15 & 4.9 & 35 & 7.1 & 56.8 & $2353^{+51}_{-63}$ & 2030 & 635 & $468.0^{+59.0}_{-53.0}$ & $4.1^{+0.2}_{-0.2}$ & $0.20^{+0.02}_{-0.02}$ & $0.19^{+0.03}_{-0.03}$ & $0.70^{+0.02}_{-0.02}$ \\
NGC 2506 & 0.002 & $2237^{+6}_{-5}$ & $3058^{+430}_{--414}$ & $-0.38^{+0.01}_{-0.01}$ & $280^{+4}_{-4}$ & 0.07 & 3.87 & 20 & 3.3 & 38.1 & $1491^{+19}_{-24}$ & 1619 & 419 & $335.0^{+47.0}_{-42.0}$ & $6.7^{+0.7}_{-0.7}$ & $0.20^{+0.02}_{-0.02}$ & $0.23^{+0.03}_{-0.03}$ & $0.67^{+0.01}_{-0.01}$ \\
NGC 6819 & 0.106 & $2395^{+3}_{-1}$ & $2435^{+717}_{--711}$ & $-0.06^{+0.00}_{-0.00}$ & $558^{+1}_{-1}$ & 0.06 & 2.54 & 24 & 2.8 & 39.0 & $1157^{+26}_{-35}$ & 1162 & 270 & $158.0^{+24.0}_{-22.0}$ & $15.2^{+4.5}_{-4.5}$ & $0.22^{+0.03}_{-0.03}$ & $0.32^{+0.06}_{-0.06}$ & $0.79^{+0.03}_{-0.03}$ \\
NGC 2420 & 0.002 & $2408^{+5}_{-4}$ & $2410^{+198}_{--189}$ & $-0.28^{+0.00}_{-0.00}$ & $168^{+2}_{-3}$ & 0.04 & 1.86 & 18 & 2.0 & 32.7 & $491^{+8}_{-10}$ & 672 & 138 & $94.0^{+20.0}_{-18.0}$ & $25.6^{+24.1}_{-24.1}$ & $0.24^{+0.03}_{-0.03}$ & $0.27^{+0.06}_{-0.06}$ & $0.82^{+0.02}_{-0.02}$ \\
Haffner 22 & 0.081 & $2497^{+6}_{-5}$ & $2518^{+726}_{--714}$ & $-0.17^{+0.00}_{-0.00}$ & $546^{+2}_{-2}$ & 0.07 & 3.01 & 41 & 2.7 & 65.8 & $312^{+7}_{-9}$ & 431 & 65 & $141.0^{+38.0}_{-33.0}$ & $17.7^{+17.2}_{-17.2}$ & $0.24^{+0.05}_{-0.05}$ & $0.24^{+0.11}_{-0.11}$ & $0.73^{+0.07}_{-0.07}$ \\
Ruprecht 171 & 0.002 & $2809^{+11}_{-7}$ & $1533^{+856}_{--847}$ & $-0.17^{+0.00}_{-0.00}$ & $960^{+2}_{-3}$ & 0.09 & 2.52 & 23 & 4.4 & 42.1 & $777^{+16}_{-18}$ & 705 & 213 & $129.0^{+27.0}_{-24.0}$ & $21.8^{+16.4}_{-16.4}$ & $0.35^{+0.03}_{-0.03}$ & $0.39^{+0.06}_{-0.06}$ & $0.76^{+0.02}_{-0.02}$ \\
Berkeley 32 & 0.275 & $3102^{+15}_{-13}$ & $3333^{+1114}_{--1092}$ & $-0.26^{+0.01}_{-0.01}$ & $620^{+7}_{-6}$ & 0.04 & 2.29 & 25 & 1.5 & 56.3 & $284^{+7}_{-10}$ & 430 & 61 & $95.0^{+28.0}_{-24.0}$ & $32.6^{+68.0}_{-68.0}$ & $0.24^{+0.05}_{-0.05}$ & $0.19^{+0.10}_{-0.10}$ & $0.72^{+0.03}_{-0.03}$ \\
NGC 2682 & 0.002 & $3771^{+6}_{-5}$ & $845^{+64}_{--62}$ & $-0.04^{+0.03}_{-0.00}$ & $156^{+2}_{-13}$ & 0.17 & 2.54 & 75 & 8.0 & 100.8 & $964^{+20}_{-28}$ & 1314 & 293 & $190.0^{+28.0}_{-25.0}$ & $19.8^{+6.8}_{-6.8}$ & $0.27^{+0.03}_{-0.03}$ & $0.34^{+0.05}_{-0.05}$ & $0.78^{+0.01}_{-0.01}$ \\
Berkeley 39 & 0.012 & $6052^{+29}_{-32}$ & $4042^{+1182}_{--1149}$ & $-0.25^{+0.01}_{-0.01}$ & $549^{+6}_{-5}$ & 0.05 & 3.59 & 21 & 2.3 & 22.5 & $667^{+16}_{-24}$ & 728 & 198 & $234.0^{+44.0}_{-39.0}$ & $25.9^{+19.2}_{-19.2}$ & $0.34^{+0.03}_{-0.03}$ & $0.50^{+0.06}_{-0.06}$ & $0.80^{+0.03}_{-0.03}$ \\
NGC 188 & 0.002 & $6126^{+62}_{-30}$ & $1818^{+275}_{--263}$ & $-0.01^{+0.01}_{-0.01}$ & $301^{+3}_{-3}$ & 0.1 & 3.08 & 30 & 4.6 & 44.9 & $907^{+19}_{-27}$ & 1017 & 268 & $218.0^{+37.0}_{-32.0}$ & $28.1^{+17.0}_{-17.0}$ & $0.29^{+0.03}_{-0.03}$ & $0.33^{+0.05}_{-0.05}$ & $0.72^{+0.03}_{-0.03}$ \\
\bottomrule \end{tabular} } 
\vspace{1em}
\begin{minipage}{\linewidth}
\footnotesize
\textbf{Notes:} \textbf{Cluster} – cluster name;
\textbf{min $p$-value} – minimum p-value;
\textbf{Age/Myr} – cluster age in Myr;
\textbf{Distance/pc} – heliocentric distance;
\textbf{[Fe/H]} – metallicity;
\textbf{$A_{\rm V}$/mmag} – extinction;
\textbf{$R_{\rm h}^{\circ}$, $R_{\rm h}$/pc} – half-mass radii in angular and physical units;
\textbf{$r_{\rm eff}^{\prime}$}, \textbf{$r_{\rm c}^{\prime}$}, \textbf{$r_{\rm t}^{\prime}$} – effective/core/tidal radii;
\textbf{$M/M_{\odot}$} – cluster mass;
\textbf{$N$} – total stars within the effective radius;
\textbf{$N_{\rm c}$} – number of stars in the core;
\textbf{$t_{\rm rh}$/Myr} – half-mass relaxation time;
\textbf{Age/$t_{\rm rh}$} – dynamical age;
\textbf{$f_{\rm{b,q,M}}$}, \textbf{$f_{\rm{b,q,M,rc}}$} – binary fractions;
\textbf{Median $q$} – median mass ratio of ML-MS binaries with $q>0.5$.
\end{minipage}
\end{table}
\end{turnpage}
\clearpage

\begin{figure*}
\includegraphics[width=\textwidth]{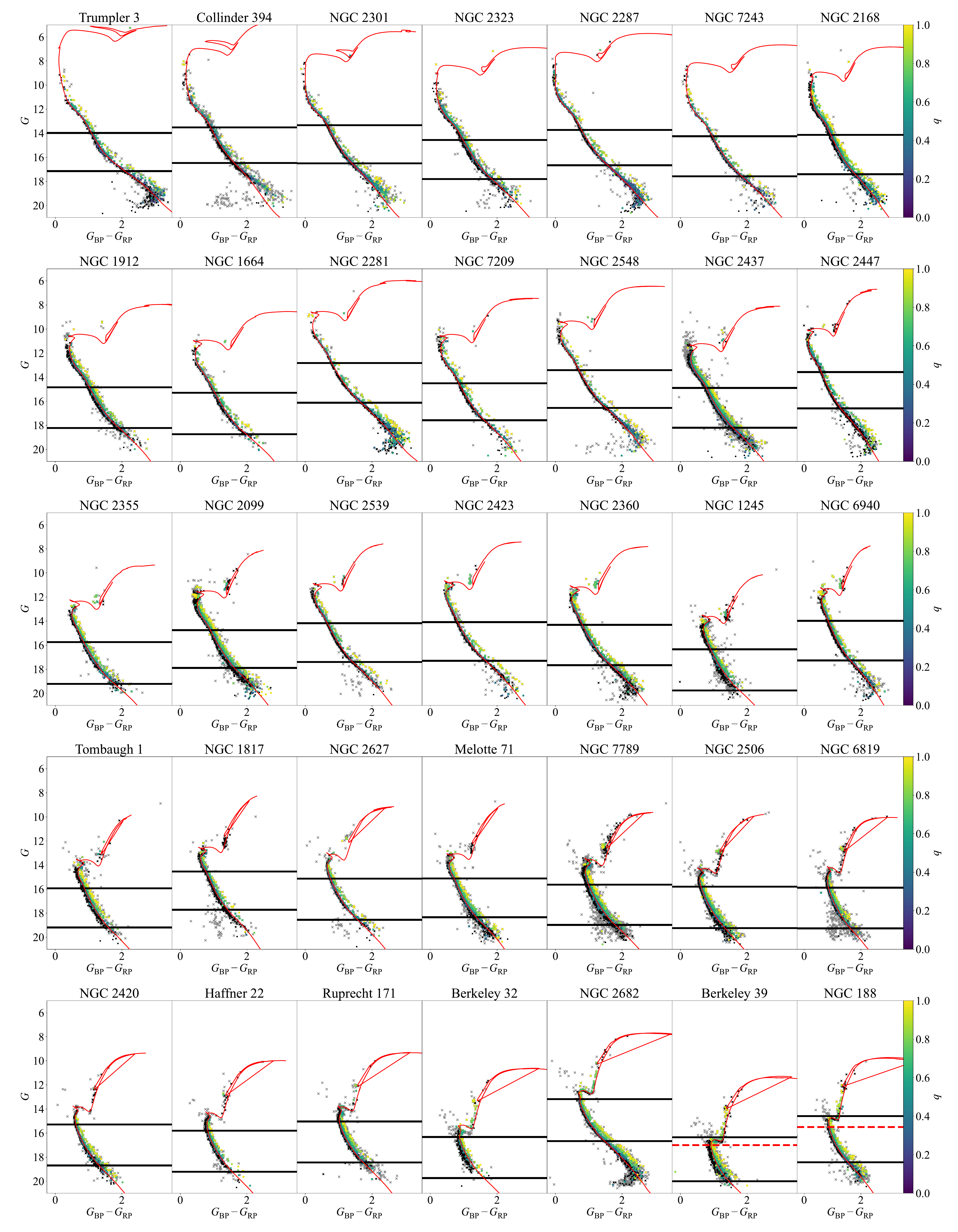}
    \caption{CMDs of all the cluster members in our full sample of 35 OCs, as well as the isochrone of best fit in red.  The stars are colored black if they are single stars and the binary stars are colored by their mass ratio.   Members from \cite{Hunt2023} that lie within the radial extent we consider are shown with gray 'X's.  The region between the black horizontal lines shows the ML-MS range for each cluster.  For two clusters, the maximum ML-MS mass is above the turnoff, and we therefore impose a lower limit shown by the red dashed lines. The clusters are shown in order of increasing cluster age.} 
    \label{fig:CMDs}
\end{figure*}

\begin{figure*}
\includegraphics[width=\textwidth]{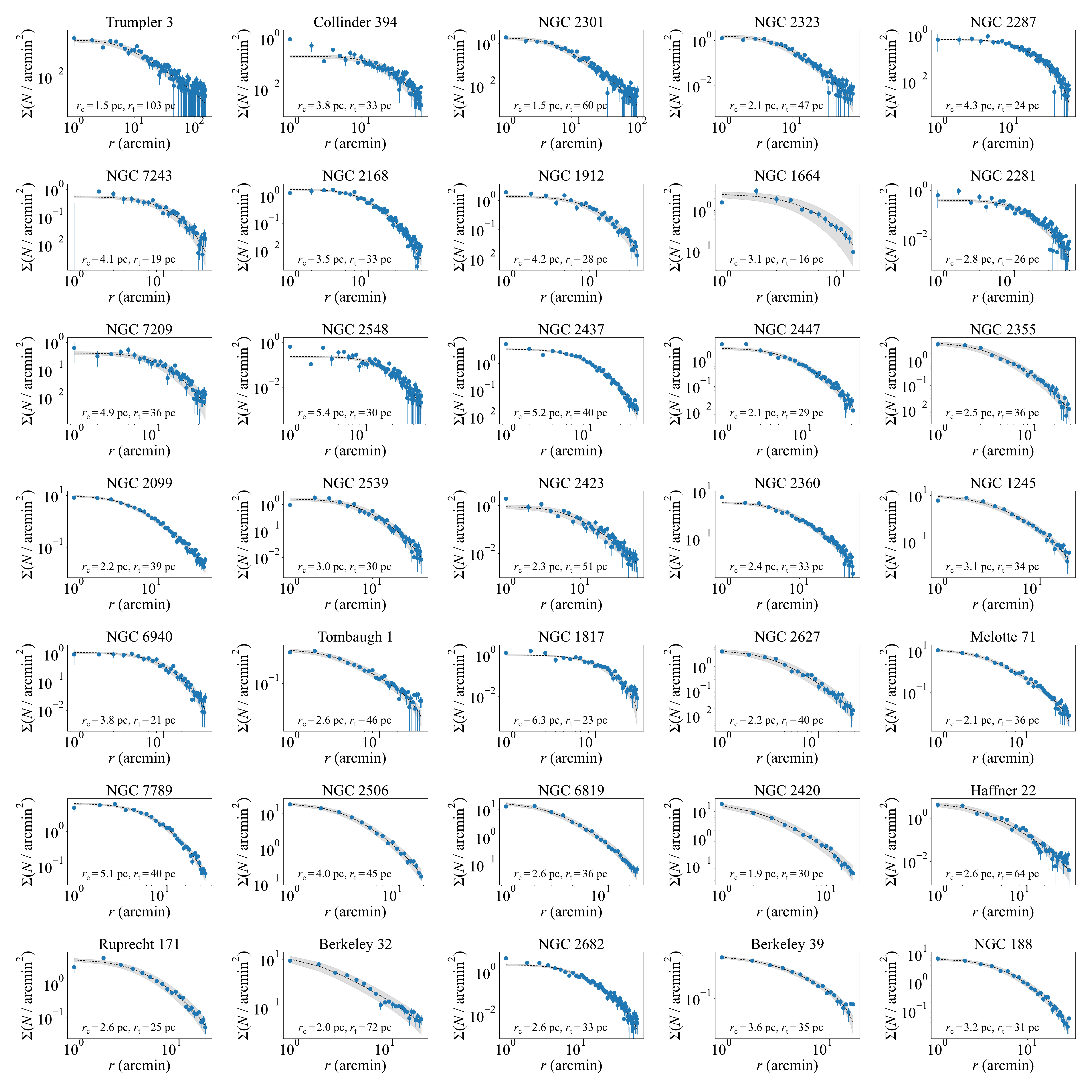}
    \caption{King model fits to the data (blue points) in our full sample (35 OCs).  The gray shaded region shows the $1-\sigma$ bounds for the fit.} 
    \label{fig:King_fits}
\end{figure*}

\begin{figure*}[hbt!]
\includegraphics[width=\columnwidth]{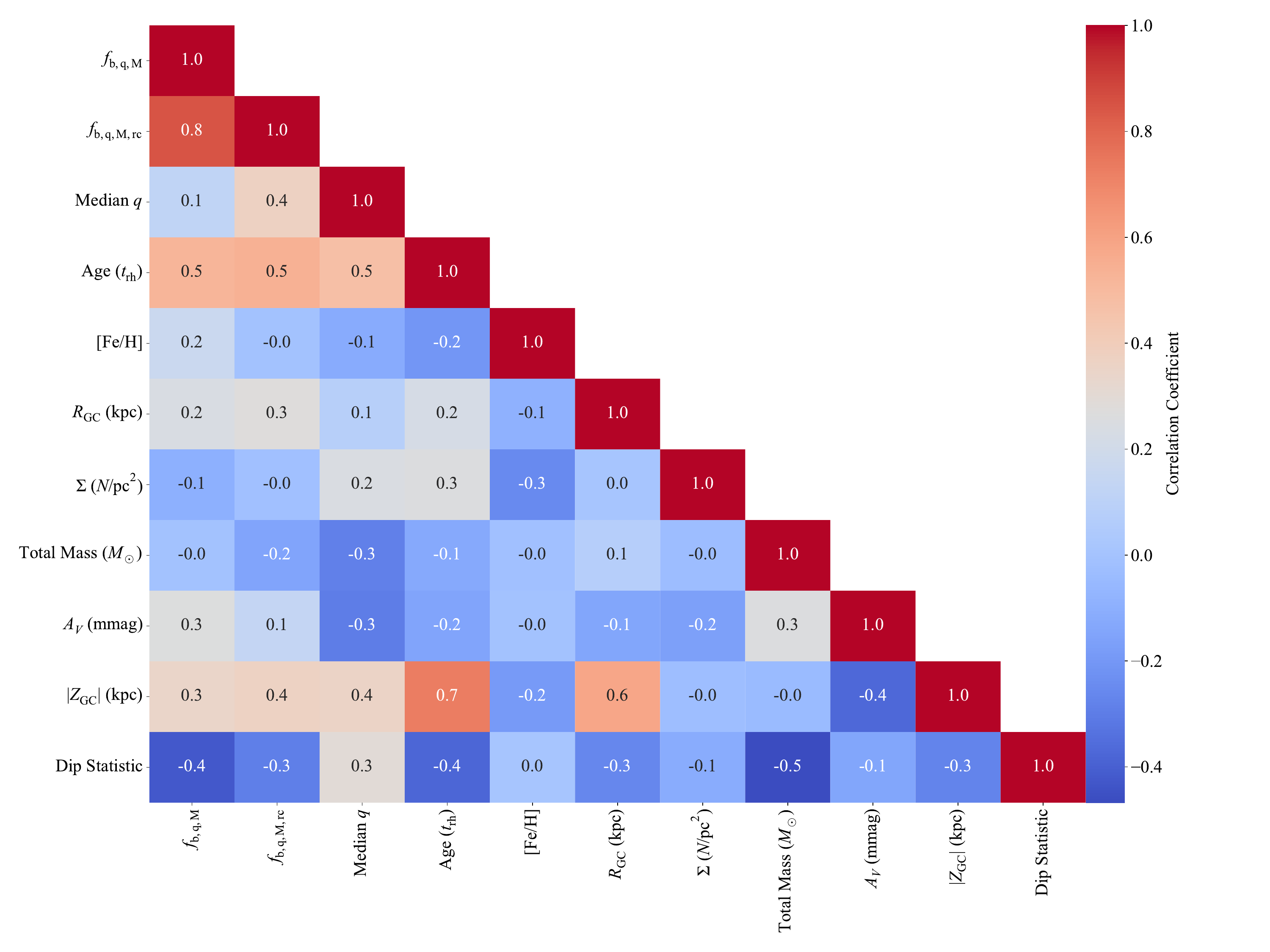}
    \caption{Correlation matrix for various cluster parameters in our primary sample of 31 OCs.} 
    \label{fig:allOC_map}
\end{figure*}

\begin{figure*}[hbt!]
\includegraphics[width=\columnwidth]{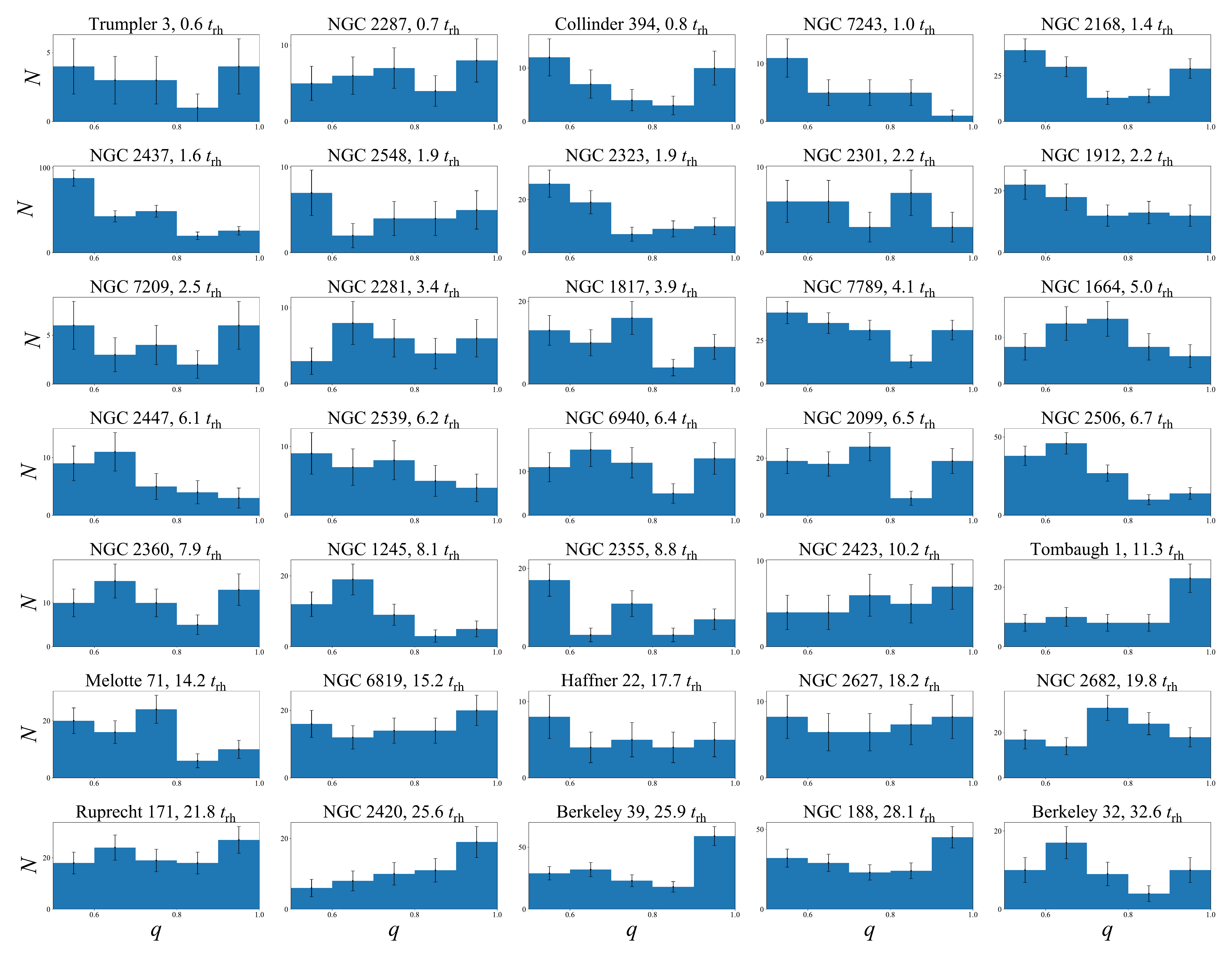}
    \caption{Mass-ratio ($q$) distributions for $q>0.5$  for all ML-MS binaries within three core radii for all 35 OCs.} 
    \label{fig:all_q_tr}
\end{figure*}

\bibliography{main}{}

\begin{thebibliography}{}
\expandafter\ifx\csname natexlab\endcsname\relax\def\natexlab#1{#1}\fi
\providecommand{\url}[1]{\href{#1}{#1}}
\providecommand{\dodoi}[1]{doi:~\href{http://doi.org/#1}{\nolinkurl{#1}}}
\providecommand{\doeprint}[1]{\href{http://ascl.net/#1}{\nolinkurl{http://ascl.net/#1}}}
\providecommand{\doarXiv}[1]{\href{https://arxiv.org/abs/#1}{\nolinkurl{https://arxiv.org/abs/#1}}}

\bibitem[{{Albrow}(2024)}]{Albrow2024}
{Albrow}, M.~D. 2024, \mnras, 528, 6211, \dodoi{10.1093/mnras/stae425}

\bibitem[{{Alexander} \& {Albrow}(2025)}]{Alexander2025}
{Alexander}, J.~S., \& {Albrow}, M.~D. 2025, \mnras, 536, 471, \dodoi{10.1093/mnras/stae2636}

\bibitem[{{Alfonso} {et~al.}(2024){Alfonso}, {Garc{\'\i}a-Varela}, \& {Vieira}}]{Alfonso2024}
{Alfonso}, J., {Garc{\'\i}a-Varela}, A., \& {Vieira}, K. 2024, \aap, 689, A18, \dodoi{10.1051/0004-6361/202450901}

\bibitem[{{Astropy Collaboration} {et~al.}(2013){Astropy Collaboration}, {Robitaille}, {Tollerud}, {Greenfield}, {Droettboom}, {Bray}, {Aldcroft}, {Davis}, {Ginsburg}, {Price-Whelan}, {Kerzendorf}, {Conley}, {Crighton}, {Barbary}, {Muna}, {Ferguson}, {Grollier}, {Parikh}, {Nair}, {Unther}, {Deil}, {Woillez}, {Conseil}, {Kramer}, {Turner}, {Singer}, {Fox}, {Weaver}, {Zabalza}, {Edwards}, {Azalee Bostroem}, {Burke}, {Casey}, {Crawford}, {Dencheva}, {Ely}, {Jenness}, {Labrie}, {Lim}, {Pierfederici}, {Pontzen}, {Ptak}, {Refsdal}, {Servillat}, \& {Streicher}}]{astropy:2013}
{Astropy Collaboration}, {Robitaille}, T.~P., {Tollerud}, E.~J., {et~al.} 2013, \aap, 558, A33, \dodoi{10.1051/0004-6361/201322068}

\bibitem[{{Astropy Collaboration} {et~al.}(2018){Astropy Collaboration}, {Price-Whelan}, {Sip{\H{o}}cz}, {G{\"u}nther}, {Lim}, {Crawford}, {Conseil}, {Shupe}, {Craig}, {Dencheva}, {Ginsburg}, {VanderPlas}, {Bradley}, {P{\'e}rez-Su{\'a}rez}, {de Val-Borro}, {Aldcroft}, {Cruz}, {Robitaille}, {Tollerud}, {Ardelean}, {Babej}, {Bach}, {Bachetti}, {Bakanov}, {Bamford}, {Barentsen}, {Barmby}, {Baumbach}, {Berry}, {Biscani}, {Boquien}, {Bostroem}, {Bouma}, {Brammer}, {Bray}, {Breytenbach}, {Buddelmeijer}, {Burke}, {Calderone}, {Cano Rodr{\'\i}guez}, {Cara}, {Cardoso}, {Cheedella}, {Copin}, {Corrales}, {Crichton}, {D'Avella}, {Deil}, {Depagne}, {Dietrich}, {Donath}, {Droettboom}, {Earl}, {Erben}, {Fabbro}, {Ferreira}, {Finethy}, {Fox}, {Garrison}, {Gibbons}, {Goldstein}, {Gommers}, {Greco}, {Greenfield}, {Groener}, {Grollier}, {Hagen}, {Hirst}, {Homeier}, {Horton}, {Hosseinzadeh}, {Hu}, {Hunkeler}, {Ivezi{\'c}}, {Jain}, {Jenness}, {Kanarek}, {Kendrew}, {Kern}, {Kerzendorf}, {Khvalko}, {King}, {Kirkby}, {Kulkarni},
  {Kumar}, {Lee}, {Lenz}, {Littlefair}, {Ma}, {Macleod}, {Mastropietro}, {McCully}, {Montagnac}, {Morris}, {Mueller}, {Mumford}, {Muna}, {Murphy}, {Nelson}, {Nguyen}, {Ninan}, {N{\"o}the}, {Ogaz}, {Oh}, {Parejko}, {Parley}, {Pascual}, {Patil}, {Patil}, {Plunkett}, {Prochaska}, {Rastogi}, {Reddy Janga}, {Sabater}, {Sakurikar}, {Seifert}, {Sherbert}, {Sherwood-Taylor}, {Shih}, {Sick}, {Silbiger}, {Singanamalla}, {Singer}, {Sladen}, {Sooley}, {Sornarajah}, {Streicher}, {Teuben}, {Thomas}, {Tremblay}, {Turner}, {Terr{\'o}n}, {van Kerkwijk}, {de la Vega}, {Watkins}, {Weaver}, {Whitmore}, {Woillez}, {Zabalza}, \& {Astropy Contributors}}]{astropy:2018}
{Astropy Collaboration}, {Price-Whelan}, A.~M., {Sip{\H{o}}cz}, B.~M., {et~al.} 2018, \aj, 156, 123, \dodoi{10.3847/1538-3881/aabc4f}

\bibitem[{{Astropy Collaboration} {et~al.}(2022){Astropy Collaboration}, {Price-Whelan}, {Lim}, {Earl}, {Starkman}, {Bradley}, {Shupe}, {Patil}, {Corrales}, {Brasseur}, {N{"o}the}, {Donath}, {Tollerud}, {Morris}, {Ginsburg}, {Vaher}, {Weaver}, {Tocknell}, {Jamieson}, {van Kerkwijk}, {Robitaille}, {Merry}, {Bachetti}, {G{"u}nther}, {Aldcroft}, {Alvarado-Montes}, {Archibald}, {B{'o}di}, {Bapat}, {Barentsen}, {Baz{'a}n}, {Biswas}, {Boquien}, {Burke}, {Cara}, {Cara}, {Conroy}, {Conseil}, {Craig}, {Cross}, {Cruz}, {D'Eugenio}, {Dencheva}, {Devillepoix}, {Dietrich}, {Eigenbrot}, {Erben}, {Ferreira}, {Foreman-Mackey}, {Fox}, {Freij}, {Garg}, {Geda}, {Glattly}, {Gondhalekar}, {Gordon}, {Grant}, {Greenfield}, {Groener}, {Guest}, {Gurovich}, {Handberg}, {Hart}, {Hatfield-Dodds}, {Homeier}, {Hosseinzadeh}, {Jenness}, {Jones}, {Joseph}, {Kalmbach}, {Karamehmetoglu}, {Ka{l}uszy{'n}ski}, {Kelley}, {Kern}, {Kerzendorf}, {Koch}, {Kulumani}, {Lee}, {Ly}, {Ma}, {MacBride}, {Maljaars}, {Muna}, {Murphy}, {Norman}, {O'Steen},
  {Oman}, {Pacifici}, {Pascual}, {Pascual-Granado}, {Patil}, {Perren}, {Pickering}, {Rastogi}, {Roulston}, {Ryan}, {Rykoff}, {Sabater}, {Sakurikar}, {Salgado}, {Sanghi}, {Saunders}, {Savchenko}, {Schwardt}, {Seifert-Eckert}, {Shih}, {Jain}, {Shukla}, {Sick}, {Simpson}, {Singanamalla}, {Singer}, {Singhal}, {Sinha}, {Sip{H{o}}cz}, {Spitler}, {Stansby}, {Streicher}, {{{S}}umak}, {Swinbank}, {Taranu}, {Tewary}, {Tremblay}, {Val-Borro}, {Van Kooten}, {Vasovi{'c}}, {Verma}, {de Miranda Cardoso}, {Williams}, {Wilson}, {Winkel}, {Wood-Vasey}, {Xue}, {Yoachim}, {Zhang}, {Zonca}, \& {Astropy Project Contributors}}]{astropy:2022}
{Astropy Collaboration}, {Price-Whelan}, A.~M., {Lim}, P.~L., {et~al.} 2022, \apj, 935, 167, \dodoi{10.3847/1538-4357/ac7c74}

\bibitem[{Bate(2000)}]{Bate2000}
Bate, M.~R. 2000, Monthly Notices of the Royal Astronomical Society, 314, 33

\bibitem[{{Bate}(2009)}]{Bate2009}
{Bate}, M.~R. 2009, \mnras, 392, 590, \dodoi{10.1111/j.1365-2966.2008.14106.x}

\bibitem[{{Bressan} {et~al.}(2012){Bressan}, {Marigo}, {Girardi}, {Salasnich}, {Dal Cero}, {Rubele}, \& {Nanni}}]{Bressan2012}
{Bressan}, A., {Marigo}, P., {Girardi}, L., {et~al.} 2012, \mnras, 427, 127, \dodoi{10.1111/j.1365-2966.2012.21948.x}

\bibitem[{{Bressert} {et~al.}(2010){Bressert}, {Bastian}, {Gutermuth}, {Megeath}, {Allen}, {Evans}, {Rebull}, {Hatchell}, {Johnstone}, {Bourke}, {Cieza}, {Harvey}, {Merin}, {Ray}, \& {Tothill}}]{bre10}
{Bressert}, E., {Bastian}, N., {Gutermuth}, R., {et~al.} 2010, Monthly Notices of the Royal Astronomical Society, 409, L54, \dodoi{10.1111/j.1745-3933.2010.00946.x}

\bibitem[{{Caballero-Nieves} {et~al.}(2014){Caballero-Nieves}, {Nelan}, {Gies}, {Wallace}, {DeGioia-Eastwood}, {Herrero}, {Jao}, {Mason}, {Massey}, {Moffat}, \& {Walborn}}]{cab14}
{Caballero-Nieves}, S.~M., {Nelan}, E.~P., {Gies}, D.~R., {et~al.} 2014, The Astronomical Journal, 147, 40, \dodoi{10.1088/0004-6256/147/2/40}

\bibitem[{{Chabrier}(2003)}]{Chabrier2003}
{Chabrier}, G. 2003, \pasp, 115, 763, \dodoi{10.1086/376392}

\bibitem[{{Childs} \& {Geller}(2025)}]{Childs_Zenodo}
{Childs}, A.~C., \& {Geller}, A.~M. 2025, \dodoi{10.5281/zenodo.15733690}

\bibitem[{{Childs} {et~al.}(2024){Childs}, {Geller}, {von Hippel}, {Motherway}, \& {Zwicker}}]{Childs2024}
{Childs}, A.~C., {Geller}, A.~M., {von Hippel}, T., {Motherway}, E., \& {Zwicker}, C. 2024, \apj, 962, 41, \dodoi{10.3847/1538-4357/ad18c0}

\bibitem[{{Cohen} {et~al.}(2020){Cohen}, {Geller}, \& {von Hippel}}]{Cohen2020}
{Cohen}, R.~E., {Geller}, A.~M., \& {von Hippel}, T. 2020, \aj, 159, 11, \dodoi{10.3847/1538-3881/ab59d7}

\bibitem[{{Cordoni} {et~al.}(2023){Cordoni}, {Milone}, {Marino}, {Vesperini}, {Dondoglio}, {Legnardi}, {Mohandasan}, {Carlos}, {Lagioia}, {Jang}, \& {Ziliotto}}]{Cordoni2023}
{Cordoni}, G., {Milone}, A.~P., {Marino}, A.~F., {et~al.} 2023, \aap, 672, A29, \dodoi{10.1051/0004-6361/202245457}

\bibitem[{{Cote} {et~al.}(1991){Cote}, {Richer}, \& {Fahlman}}]{Cote1991}
{Cote}, P., {Richer}, H.~B., \& {Fahlman}, G.~G. 1991, \aj, 102, 1358, \dodoi{10.1086/115961}

\bibitem[{{de La Fuente Marcos}(1998)}]{delaFuente1998}
{de La Fuente Marcos}, R. 1998, \pasp, 110, 1117, \dodoi{10.1086/316220}

\bibitem[{{Dias} {et~al.}(2021){Dias}, {Monteiro}, {Moitinho}, {L{\'e}pine}, {Carraro}, {Paunzen}, {Alessi}, \& {Villela}}]{Dias2021}
{Dias}, W.~S., {Monteiro}, H., {Moitinho}, A., {et~al.} 2021, \mnras, 504, 356, \dodoi{10.1093/mnras/stab770}

\bibitem[{{Donada} {et~al.}(2023){Donada}, {Anders}, {Jordi}, {Masana}, {Gieles}, {Perren}, {Balaguer-N{\'u}{\~n}ez}, {Castro-Ginard}, {Cantat-Gaudin}, \& {Casamiquela}}]{Donada2023}
{Donada}, J., {Anders}, F., {Jordi}, C., {et~al.} 2023, \aap, 675, A89, \dodoi{10.1051/0004-6361/202245219}

\bibitem[{{Duch{\^e}ne} \& {Kraus}(2013)}]{Duchene2013}
{Duch{\^e}ne}, G., \& {Kraus}, A. 2013, \araa, 51, 269, \dodoi{10.1146/annurev-astro-081710-102602}

\bibitem[{{Evans} {et~al.}(2009){Evans}, {Dunham}, {J{\o}rgensen}, {Enoch}, {Mer{\'\i}n}, {van Dishoeck}, {Alcal{\'a}}, {Myers}, {Stapelfeldt}, {Huard}, {Allen}, {Harvey}, {van Kempen}, {Blake}, {Koerner}, {Mundy}, {Padgett}, \& {Sargent}}]{eva09}
{Evans}, Neal~J., I., {Dunham}, M.~M., {J{\o}rgensen}, J.~K., {et~al.} 2009, The Astrophysical Journal Supplement Series, 181, 321, \dodoi{10.1088/0067-0049/181/2/321}

\bibitem[{{Fregeau} {et~al.}(2009){Fregeau}, {Ivanova}, \& {Rasio}}]{Fregeau2009}
{Fregeau}, J.~M., {Ivanova}, N., \& {Rasio}, F.~A. 2009, \apj, 707, 1533, \dodoi{10.1088/0004-637X/707/2/1533}

\bibitem[{Fregeau {et~al.}(2002)Fregeau, Joshi, Zwart, \& Rasio}]{Fregeau_2002}
Fregeau, J.~M., Joshi, K.~J., Zwart, S. F.~P., \& Rasio, F.~A. 2002, The Astrophysical Journal, 570, 171

\bibitem[{{Friel}(1995)}]{Friel1995}
{Friel}, E.~D. 1995, \araa, 33, 381, \dodoi{10.1146/annurev.aa.33.090195.002121}

\bibitem[{{Frinchaboy} \& {Thompson}(2015)}]{Frinchaboy2015}
{Frinchaboy}, P., \& {Thompson}, B. 2015, arXiv e-prints, arXiv:1511.04789.
\newblock \doarXiv{1511.04789}

\bibitem[{Geller {et~al.}(2012)Geller, Hurley, \& Mathieu}]{Geller_2013}
Geller, A.~M., Hurley, J.~R., \& Mathieu, R.~D. 2012, The Astronomical Journal, 145, 8

\bibitem[{{Giersz} \& {Heggie}(1994)}]{Giersz1994}
{Giersz}, M., \& {Heggie}, D.~C. 1994, \mnras, 268, 257, \dodoi{10.1093/mnras/268.1.257}

\bibitem[{{Ginat} \& {Perets}(2024)}]{Ginat2024}
{Ginat}, Y.~B., \& {Perets}, H.~B. 2024, \mnras, 531, 739, \dodoi{10.1093/mnras/stae1241}

\bibitem[{{Goodwin}(2013)}]{Goodwin2013}
{Goodwin}, S.~P. 2013, \mnras, 430, L6, \dodoi{10.1093/mnrasl/sls037}

\bibitem[{Hartigan \& Hartigan(1985)}]{Hartigan_1985}
Hartigan, J.~A., \& Hartigan, P.~M. 1985, The Annals of Statistics, 13, 70

\bibitem[{{Heggie}(1975)}]{heggie75}
{Heggie}, D.~C. 1975, Monthly Notices of the Royal Astronomical Society, 173, 729, \dodoi{10.1093/mnras/173.3.729}

\bibitem[{{Heggie}(2003)}]{Heggie2003}
{Heggie}, D.~C. 2003, in Astrophysical Supercomputing using Particle Simulations, ed. J.~{Makino} \& P.~{Hut}, Vol. 208, 81, \dodoi{10.48550/arXiv.astro-ph/0111045}

\bibitem[{{Heiter} {et~al.}(2014){Heiter}, {Soubiran}, {Netopil}, \& {Paunzen}}]{Heiter2014}
{Heiter}, U., {Soubiran}, C., {Netopil}, M., \& {Paunzen}, E. 2014, \aap, 561, A93, \dodoi{10.1051/0004-6361/201322559}

\bibitem[{{Hunt} \& {Reffert}(2021)}]{Hunt2021}
{Hunt}, E.~L., \& {Reffert}, S. 2021, \aap, 646, A104, \dodoi{10.1051/0004-6361/202039341}

\bibitem[{{Hunt} \& {Reffert}(2023)}]{Hunt2023}
---. 2023, \aap, 673, A114, \dodoi{10.1051/0004-6361/202346285}

\bibitem[{{Hurley} \& {Tout}(1998)}]{Hurley1998}
{Hurley}, J., \& {Tout}, C.~A. 1998, \mnras, 300, 977, \dodoi{10.1046/j.1365-8711.1998.01981.x}

\bibitem[{{Jadhav} {et~al.}(2021){Jadhav}, {Roy}, {Joshi}, \& {Subramaniam}}]{Jadhav2021}
{Jadhav}, V.~V., {Roy}, K., {Joshi}, N., \& {Subramaniam}, A. 2021, \aj, 162, 264, \dodoi{10.3847/1538-3881/ac2571}

\bibitem[{{Janes} \& {Phelps}(1994)}]{Janes1994}
{Janes}, K.~A., \& {Phelps}, R.~L. 1994, \aj, 108, 1773, \dodoi{10.1086/117192}

\bibitem[{{Ji} \& {Bregman}(2015)}]{Ji2015}
{Ji}, J., \& {Bregman}, J.~N. 2015, \apj, 807, 32, \dodoi{10.1088/0004-637X/807/1/32}

\bibitem[{{Jiang} {et~al.}(2024){Jiang}, {Zhong}, {Qin}, {Tang}, {Chen}, \& {Hou}}]{Jiang2024}
{Jiang}, Y., {Zhong}, J., {Qin}, S., {et~al.} 2024, \apj, 971, 71, \dodoi{10.3847/1538-4357/ad5344}

\bibitem[{{King}(1962)}]{King1962}
{King}, I. 1962, \aj, 67, 471, \dodoi{10.1086/108756}

\bibitem[{{Lada} \& {Lada}(2003)}]{lad03}
{Lada}, C.~J., \& {Lada}, E.~A. 2003, Annual Review of Astronomy and Astrophysics, 41, 57, \dodoi{10.1146/annurev.astro.41.011802.094844}

\bibitem[{Layden {et~al.}(1999)Layden, Ritter, Welch, \& Webb}]{Layden_1999}
Layden, A.~C., Ritter, L.~A., Welch, D.~L., \& Webb, T. M.~A. 1999, The Astronomical Journal, 117, 1313

\bibitem[{{Leiner} {et~al.}(2019){Leiner}, {Mathieu}, {Vanderburg}, {Gosnell}, \& {Smith}}]{Leiner2019}
{Leiner}, E., {Mathieu}, R.~D., {Vanderburg}, A., {Gosnell}, N.~M., \& {Smith}, J.~C. 2019, \apj, 881, 47, \dodoi{10.3847/1538-4357/ab2bf8}

\bibitem[{{Li} {et~al.}(2020){Li}, {Shao}, {Li}, {Yu}, {Zhong}, \& {Chen}}]{Li2020}
{Li}, L., {Shao}, Z., {Li}, Z.-Z., {et~al.} 2020, \apj, 901, 49, \dodoi{10.3847/1538-4357/abaef3}

\bibitem[{{Lindegren} {et~al.}(2021){Lindegren}, {Bastian}, {Biermann}, {Bombrun}, {de Torres}, {Gerlach}, {Geyer}, {Hern{\'a}ndez}, {Hilger}, {Hobbs}, {Klioner}, {Lammers}, {McMillan}, {Ramos-Lerate}, {Steidelm{\"u}ller}, {Stephenson}, \& {van Leeuwen}}]{Lindegren2021}
{Lindegren}, L., {Bastian}, U., {Biermann}, M., {et~al.} 2021, \aap, 649, A4, \dodoi{10.1051/0004-6361/202039653}

\bibitem[{{Liu} {et~al.}(2025{\natexlab{a}}){Liu}, {Shao}, \& {Li}}]{Liu2025}
{Liu}, R., {Shao}, Z., \& {Li}, L. 2025{\natexlab{a}}, \aj, 169, 116, \dodoi{10.3847/1538-3881/ada380}

\bibitem[{{Liu} {et~al.}(2025{\natexlab{b}}){Liu}, {Shao}, \& {Li}}]{Liu2025b}
---. 2025{\natexlab{b}}, \aj, 169, 116, \dodoi{10.3847/1538-3881/ada380}

\bibitem[{{Majewski} {et~al.}(2017){Majewski}, {Schiavon}, {Frinchaboy}, {Allende Prieto}, {Barkhouser}, {Bizyaev}, {Blank}, {Brunner}, {Burton}, {Carrera}, {Chojnowski}, {Cunha}, {Epstein}, {Fitzgerald}, {Garc{\'\i}a P{\'e}rez}, {Hearty}, {Henderson}, {Holtzman}, {Johnson}, {Lam}, {Lawler}, {Maseman}, {M{\'e}sz{\'a}ros}, {Nelson}, {Nguyen}, {Nidever}, {Pinsonneault}, {Shetrone}, {Smee}, {Smith}, {Stolberg}, {Skrutskie}, {Walker}, {Wilson}, {Zasowski}, {Anders}, {Basu}, {Beland}, {Blanton}, {Bovy}, {Brownstein}, {Carlberg}, {Chaplin}, {Chiappini}, {Eisenstein}, {Elsworth}, {Feuillet}, {Fleming}, {Galbraith-Frew}, {Garc{\'\i}a}, {Garc{\'\i}a-Hern{\'a}ndez}, {Gillespie}, {Girardi}, {Gunn}, {Hasselquist}, {Hayden}, {Hekker}, {Ivans}, {Kinemuchi}, {Klaene}, {Mahadevan}, {Mathur}, {Mosser}, {Muna}, {Munn}, {Nichol}, {O'Connell}, {Parejko}, {Robin}, {Rocha-Pinto}, {Schultheis}, {Serenelli}, {Shane}, {Silva Aguirre}, {Sobeck}, {Thompson}, {Troup}, {Weinberg}, \& {Zamora}}]{Majewski_2017}
{Majewski}, S.~R., {Schiavon}, R.~P., {Frinchaboy}, P.~M., {et~al.} 2017, \aj, 154, 94, \dodoi{10.3847/1538-3881/aa784d}

\bibitem[{{Malofeeva} {et~al.}(2023){Malofeeva}, {Mikhnevich}, {Carraro}, \& {Seleznev}}]{Malofeeva2023}
{Malofeeva}, A.~A., {Mikhnevich}, V.~O., {Carraro}, G., \& {Seleznev}, A.~F. 2023, \aj, 165, 45, \dodoi{10.3847/1538-3881/aca666}

\bibitem[{{Malofeeva} {et~al.}(2022){Malofeeva}, {Seleznev}, \& {Carraro}}]{Malofeeva2022}
{Malofeeva}, A.~A., {Seleznev}, A.~F., \& {Carraro}, G. 2022, \aj, 163, 113, \dodoi{10.3847/1538-3881/ac47a3}

\bibitem[{{Mansbach}(1970)}]{Mansbach1970}
{Mansbach}, P. 1970, \apj, 160, 135, \dodoi{10.1086/150412}

\bibitem[{{Marks} {et~al.}(2011){Marks}, {Kroupa}, \& {Oh}}]{marks2011}
{Marks}, M., {Kroupa}, P., \& {Oh}, S. 2011, Monthly Notices of the Royal Astronomical Society, 417, 1684, \dodoi{10.1111/j.1365-2966.2011.19257.x}

\bibitem[{{Martins} \& {Palacios}(2013)}]{Martins2013}
{Martins}, F., \& {Palacios}, A. 2013, \aap, 560, A16, \dodoi{10.1051/0004-6361/201322480}

\bibitem[{Mathieu \& Geller(2009)}]{Mathieu2009}
Mathieu, R.~D., \& Geller, A.~M. 2009, Nature, 462, 1032

\bibitem[{McInnes {et~al.}(2017)McInnes, Healy, \& Astels}]{McInnes2017}
McInnes, L., Healy, J., \& Astels, S. 2017, Journal of Open Source Software, 2, 205, \dodoi{10.21105/joss.00205}

\bibitem[{{Mikhnevich} \& {Seleznev}(2024)}]{Mikhnevich2024}
{Mikhnevich}, V.~O., \& {Seleznev}, A.~F. 2024, Astronomy Reports, 68, 121, \dodoi{10.1134/S1063772924700161}

\bibitem[{{Milone} {et~al.}(2012){Milone}, {Piotto}, {Bedin}, {Aparicio}, {Anderson}, {Sarajedini}, {Marino}, {Moretti}, {Davies}, {Chaboyer}, {Dotter}, {Hempel}, {Mar{\'\i}n-Franch}, {Majewski}, {Paust}, {Reid}, {Rosenberg}, \& {Siegel}}]{Milone2012}
{Milone}, A.~P., {Piotto}, G., {Bedin}, L.~R., {et~al.} 2012, \aap, 540, A16, \dodoi{10.1051/0004-6361/201016384}

\bibitem[{{Mohandasan} {et~al.}(2024){Mohandasan}, {Milone}, {Cordoni}, {Dondoglio}, {Lagioia}, {Legnardi}, {Ziliotto}, {Jang}, {Marino}, \& {Carlos}}]{Mohandasan2024}
{Mohandasan}, A., {Milone}, A.~P., {Cordoni}, G., {et~al.} 2024, \aap, 681, A42, \dodoi{10.1051/0004-6361/202347424}

\bibitem[{{Motherway} {et~al.}(2024){Motherway}, {Geller}, {Childs}, {Zwicker}, \& {von Hippel}}]{Motherway2024}
{Motherway}, E., {Geller}, A.~M., {Childs}, A.~C., {Zwicker}, C., \& {von Hippel}, T. 2024, \apjl, 962, L9, \dodoi{10.3847/2041-8213/ad18bf}

\bibitem[{{Offner} {et~al.}(2023){Offner}, {Moe}, {Kratter}, {Sadavoy}, {Jensen}, \& {Tobin}}]{Offner2023}
{Offner}, S.~S.~R., {Moe}, M., {Kratter}, K.~M., {et~al.} 2023, in Astronomical Society of the Pacific Conference Series, Vol. 534, Protostars and Planets VII, ed. S.~{Inutsuka}, Y.~{Aikawa}, T.~{Muto}, K.~{Tomida}, \& M.~{Tamura}, 275, \dodoi{10.48550/arXiv.2203.10066}

\bibitem[{{Pang} {et~al.}(2023){Pang}, {Wang}, {Tang}, {Rui}, {Bai}, {Li}, {Feng}, {Kouwenhoven}, {Chen}, \& {Chuang}}]{Pang2023}
{Pang}, X., {Wang}, Y., {Tang}, S.-Y., {et~al.} 2023, \aj, 166, 110, \dodoi{10.3847/1538-3881/ace76c}

\bibitem[{{Qin} {et~al.}(2023){Qin}, {Zhong}, {Tang}, \& {Chen}}]{Qin2023}
{Qin}, S., {Zhong}, J., {Tang}, T., \& {Chen}, L. 2023, \apjs, 265, 12, \dodoi{10.3847/1538-4365/acadd6}

\bibitem[{{Raghavan} {et~al.}(2010){Raghavan}, {McAlister}, {Henry}, {Latham}, {Marcy}, {Mason}, {Gies}, {White}, \& {ten Brummelaar}}]{Raghavan2010}
{Raghavan}, D., {McAlister}, H.~A., {Henry}, T.~J., {et~al.} 2010, \apjs, 190, 1, \dodoi{10.1088/0067-0049/190/1/1}

\bibitem[{{Robinson} {et~al.}(2016){Robinson}, {von Hippel}, {Stein}, {Stenning}, {Wagner-Kaiser}, {Si}, \& {van Dyk}}]{Robinson2016}
{Robinson}, E., {von Hippel}, T., {Stein}, N., {et~al.} 2016, {BASE-9: Bayesian Analysis for Stellar Evolution with nine variables}, Astrophysics Source Code Library, record ascl:1608.007.
\newblock \doeprint{1608.007}

\bibitem[{{Sana} {et~al.}(2012){Sana}, {de Mink}, {de Koter}, {Langer}, {Evans}, {Gieles}, {Gosset}, {Izzard}, {Le Bouquin}, \& {Schneider}}]{san12}
{Sana}, H., {de Mink}, S.~E., {de Koter}, A., {et~al.} 2012, Science, 337, 444, \dodoi{10.1126/science.1223344}

\bibitem[{{Skrutskie} {et~al.}(2006){Skrutskie}, {Cutri}, {Stiening}, {Weinberg}, {Schneider}, {Carpenter}, {Beichman}, {Capps}, {Chester}, {Elias}, {Huchra}, {Liebert}, {Lonsdale}, {Monet}, {Price}, {Seitzer}, {Jarrett}, {Kirkpatrick}, {Gizis}, {Howard}, {Evans}, {Fowler}, {Fullmer}, {Hurt}, {Light}, {Kopan}, {Marsh}, {McCallon}, {Tam}, {Van Dyk}, \& {Wheelock}}]{Skrutskie2006}
{Skrutskie}, M.~F., {Cutri}, R.~M., {Stiening}, R., {et~al.} 2006, \aj, 131, 1163, \dodoi{10.1086/498708}

\bibitem[{{Sollima} {et~al.}(2007){Sollima}, {Beccari}, {Ferraro}, {Fusi Pecci}, \& {Sarajedini}}]{Sollima2007}
{Sollima}, A., {Beccari}, G., {Ferraro}, F.~R., {Fusi Pecci}, F., \& {Sarajedini}, A. 2007, \mnras, 380, 781, \dodoi{10.1111/j.1365-2966.2007.12116.x}

\bibitem[{{Spangler}(2025)}]{Spangler2025RNAASS}
{Spangler}, S.~R. 2025, Research Notes of the American Astronomical Society, 9, 34, \dodoi{10.3847/2515-5172/adb48a}

\bibitem[{{Tarricq} {et~al.}(2022){Tarricq}, {Soubiran}, {Casamiquela}, {Castro-Ginard}, {Olivares}, {Miret-Roig}, \& {Galli}}]{Tarricq2022}
{Tarricq}, Y., {Soubiran}, C., {Casamiquela}, L., {et~al.} 2022, \aap, 659, A59, \dodoi{10.1051/0004-6361/202142186}

\bibitem[{{Tokovinin}(2014)}]{Tokovinin2014}
{Tokovinin}, A. 2014, \aj, 147, 87, \dodoi{10.1088/0004-6256/147/4/87}

\bibitem[{{van Dyk} {et~al.}(2009){van Dyk}, {Degennaro}, {Stein}, {Jefferys}, \& {von Hippel}}]{vanDyke2009}
{van Dyk}, D.~A., {Degennaro}, S., {Stein}, N., {Jefferys}, W.~H., \& {von Hippel}, T. 2009, Annals of Applied Statistics, 3, 117, \dodoi{10.1214/08-AOAS219SUPP}

\bibitem[{{von Hippel} {et~al.}(2006){von Hippel}, {Jefferys}, {Scott}, {Stein}, {Winget}, {De Gennaro}, {Dam}, \& {Jeffery}}]{vonHippel2006}
{von Hippel}, T., {Jefferys}, W.~H., {Scott}, J., {et~al.} 2006, \apj, 645, 1436, \dodoi{10.1086/504369}

\bibitem[{{Winters} {et~al.}(2019){Winters}, {Henry}, {Jao}, {Subasavage}, {Chatelain}, {Slatten}, {Riedel}, {Silverstein}, \& {Payne}}]{Winters2019}
{Winters}, J.~G., {Henry}, T.~J., {Jao}, W.-C., {et~al.} 2019, \aj, 157, 216, \dodoi{10.3847/1538-3881/ab05dc}

\bibitem[{{Zwicker} {et~al.}(2024){Zwicker}, {Geller}, {Childs}, {Motherway}, \& {von Hippel}}]{Zwicker2024}
{Zwicker}, C., {Geller}, A.~M., {Childs}, A.~C., {Motherway}, E., \& {von Hippel}, T. 2024, \apj, 967, 44, \dodoi{10.3847/1538-4357/ad39c6}

\end{thebibliography}
\bibliographystyle{aasjournal}



\end{document}